\documentclass[twocolumn]{aastex6}

 \pdfoutput=1

\usepackage{verbatim}
\usepackage{graphicx}
\usepackage{cleveref}
\usepackage{hyperref}
\usepackage{multirow}
\usepackage[normalem]{ulem}

\usepackage{amsmath}
\usepackage{xcolor}

\newcommand{\rr}[1]{$r_{#1}$}
\newcommand{\M}[1]{$M_{\mathrm{#1}}$}
\newcommand{\A}{\textit{A}}
\newcommand{\N}{\textit{N}}
\newcommand{\Pf}{$P(``F"|F)$}
\newcommand{\Pn}{$P(``N"|N)$}
\newcommand{\p}{$p_0$}
\newcommand{\tf}{$t_F$}
\newcommand{\tn}{$t_N$}
\newcommand{\feat}{`Featured'}
\newcommand{\notfeat}{`Not'}
\newcommand{\raw}{GZ2$_{\text{raw}}$}
\newcommand{\weighted}{GZ2$_{\text{weighted}}$}
\newcommand{\deb}{GZ2$_{\text{debiased}}$}
\newcommand{\pmachine}{$p_{\mathrm{machine}}$}
\newcommand{\ffeat}{$f_{\mathrm{featured}}$}
\newcommand{\fsmooth}{$f_{\mathrm{smooth}}$}
\newcommand{\fstar}{$f_{\mathrm{artifact}}$}

\shorttitle{Human and machine morphology classifications}
\shortauthors{Beck et al.}

\begin{document}

\title{Integrating human and machine intelligence in galaxy morphology \\classification tasks}



\author{Melanie R. Beck\altaffilmark{1}, Claudia Scarlata\altaffilmark{1}, Lucy F. Fortson\altaffilmark{1}}
\author{Chris J. Lintott\altaffilmark{2, 3}}
\author{B. D. Simmons\altaffilmark{2,4,7}}
\author{Melanie A. Galloway\altaffilmark{1}, Kyle W. Willett\altaffilmark{1}}
\author{Hugh Dickinson\altaffilmark{1}}
\author{Karen L. Masters\altaffilmark{5}}
\author{Philip J. Marshall\altaffilmark{6}}
\author{Darryl Wright\altaffilmark{2}}


\altaffiltext{1}{Minnesota Institute for Astrophysics, University of Minnesota, Minneapolis, MN 55455, USA; beck@astro.umn.edu}
\altaffiltext{2}{Oxford Astrophysics, Denys Wilkinson Building, Keble Road, Oxford OX1 3RH, UK}
\altaffiltext{3}{New College, Oxford OX1 3BN, UK}
\altaffiltext{4}{Center for Astrophysics and Space Sciences, Department of Physics, University of California, San Diego, CA 92093, USA}
\altaffiltext{5}{Institute of Cosmology and Gravitation, University of Portsmouth, Portsmouth, UK}
\altaffiltext{6}{Kavli Institute for Particle Astrophysics and Cosmology, P.O. Box 20450, MS29, Stanford, CA 94309, U.S.A.}
\altaffiltext{7}{Einstein Fellow}

\begin{abstract}
Quantifying galaxy morphology is a challenging yet scientifically rewarding task. As the scale of data continues to increase with upcoming surveys, traditional classification methods will struggle to handle the load. We present a solution through an integration of visual and automated classifications, preserving the best features of both human and machine. We demonstrate the effectiveness of such a system through a re-analysis of visual galaxy morphology classifications collected during the Galaxy Zoo 2 (GZ2) project. We reprocess the top-level question of the GZ2 decision tree with a Bayesian classification aggregation algorithm dubbed SWAP, originally developed for the Space Warps gravitational lens project. Through a simple binary classification scheme we increase the classification rate nearly 5-fold classifying 226,124 galaxies in 92 days of GZ2 project time while reproducing labels derived from GZ2 classification data with 95.7\% accuracy.

We next combine this with a Random Forest machine learning algorithm that learns on a suite of non-parametric morphology indicators widely used for automated morphologies. We develop a decision engine that delegates tasks between human and machine and demonstrate that the combined system provides at least a factor of 8 increase in the classification rate, classifying \replaced{210,543}{210,803} galaxies in just 32 days of GZ2 project time with \replaced{93.5\%}{93.1\%} accuracy. As the Random Forest algorithm requires a minimal amount of computational cost, this result has important implications for galaxy morphology identification tasks in the era of \textit{Euclid} and other large-scale surveys.
\end{abstract}


\section{Introduction} 
\label{sec:intro}

Astronomers have made use of visual galaxy morphologies to understand the dynamical structure of these systems for nearly ninety years 
\citep[e.g.,][]{Hubble1936, 
			deVauc1959, 
			Sandage1961, 
			vandenBergh1976, 
			NairAbraham2010, 
			Baillard2011}. 
The division between early-type and late-type systems corresponds, for example, to a wide range of parameters from mass and luminosity, to environment, colour, and star formation history 
\citep[e.g.,][]{Kormendy1977,  
			Dressler1980, 
			Strateva2001, 
			Blanton2003, 
			Kauffman2003, 
			Nakamura2003, 
			Shen2003, 
			Peng2010}; 
while detailed observations of morphological features such as bars and bulges 
provide information about the history of their host systems 
\citep[e.g., reviews by][]{KK04, 
			Elmegreen2008, 
			Sheth2008, 
			Masters2010, 
			Simmons2014}. 
Modern studies of morphology  divide systems into broad classes 
\citep[e.g.,][]{Conselice2006, 
			Lintott2008, 
			Kartaltepe2015, 
			Peth2016}, 
but a wealth of information can be gained from identifying new and often rare classes, such as low redshift clumpy galaxies \citep[e.g.,][]{Elmegreen2013}, polar-ring galaxies \citep[e.g.,][]{Whitmore1990}, and the green peas \citep{Cardamone2009}.

While the Galaxy Zoo project has provided a solution that scales visual classification for current surveys  by harnessing the combined power of thousands of volunteers \citep{Lintott2008, Lintott2011, Willett2013, Willett2017, Simmons2017}, producing a prolific amount of scientific output \citep[e.g.,][]{Land2008, Bamford2009, Darg2010, Schawinski2014, Galloway2015, Smethurst2016}; upcoming surveys such as~\textit{LSST} and \textit{Euclid} will require a different approach, imaging more than a billion new galaxies  \citep{LSST, Euclid}.  If detailed morphologies can be extracted for just 0.1\% of this imaging, we will have millions of images to contend with. A project of this magnitude would take more than sixty years to classify at Galaxy Zoo's current rate and configuration. Standard visual morphology methods will thus be unable to cope with the scale of data. 

Another approach has been the automated extraction of morphologies with the development of parametric \citep{Sersic1968, Odewahn2002, Peng2002}, and non-parametric \citep{Abraham1994, Conselice2003, Abraham2003, Lotz2004,
 Freeman2013} structural indicators. While these scale well to large samples 
\citep[e.g.,][]{Simard2011, 
			Griffith2012, 
			Casteels2014, 
			Holwerda2014, 
			Meert2016}, 
they often fail to capture detailed structure and can provide only statistical morphologies with large uncertainties \cite[e.g.,][]{Abraham1996, Bershady2000}.

\begin{figure*}[ht!]
\plotone{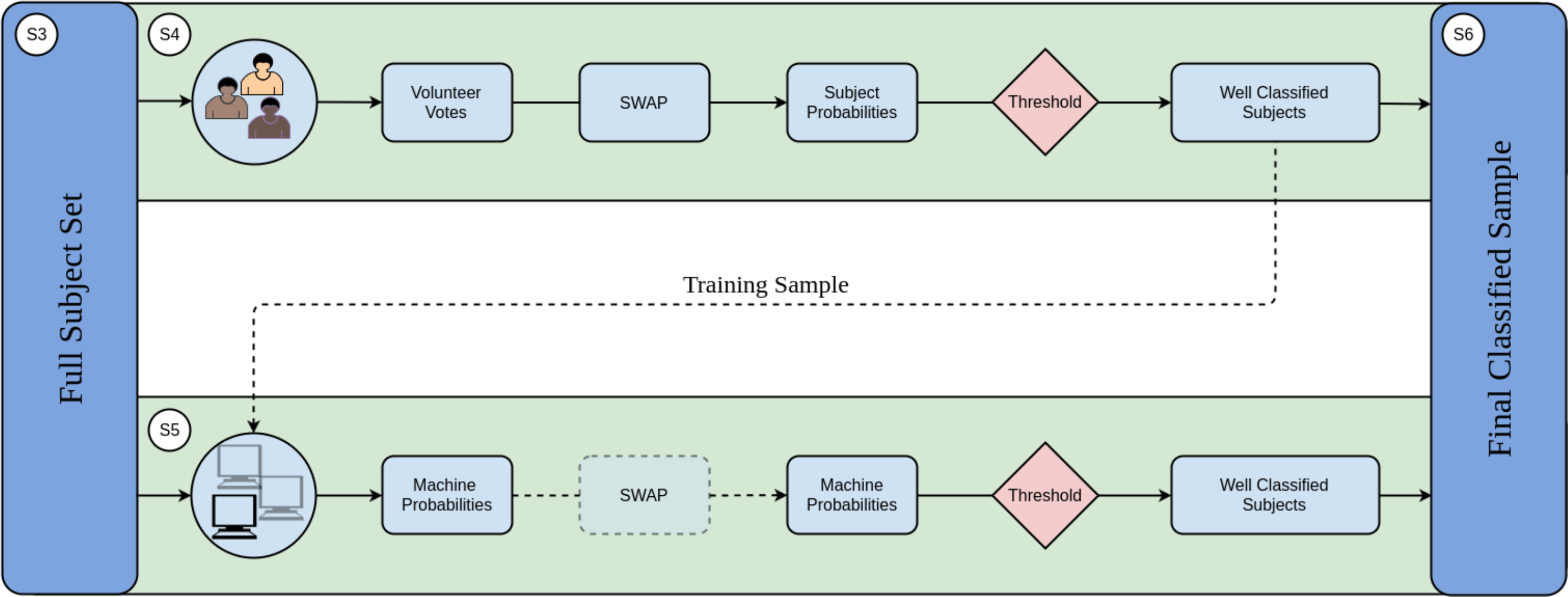}
\caption{Schematic of our hybrid system. Humans provide classifications of galaxy images via a web interface. We simulate this with the Galaxy Zoo 2 classification data described in Section~\ref{sec: data}. Human classifications are processed with an algorithm described in Section~\ref{sec: SWAP}. Subjects that pass a set of thresholds are considered human-retired (fully classified) and provide the training sample for the machine classifier as described in Section~\ref{sec: machine}. The trained machine is applied to all subjects not yet retired. Those that pass an analogous set of machine-specific thresholds are considered machine-retired. The rest remain in the system to be classified by either human or machine. This procedure is repeated  nightly. Our results are reported in Section~\ref{sec: results}.  \label{fig: schematic}}
\end{figure*}

Machine learning techniques are becoming increasingly popular for classification and image processing tasks. Another automated approach, these generally work by defining a set of features that describe the morphology in an $N$-dimensional space. The location in this morphology space defines a morphological type for each galaxy. Learning the morphology space can be achieved through algorithms such as Support Vector Machines \citep{HuertasCompany2008} or Principal Component Analysis \citep{Watanabe1985, Scarlata2007}. Another approach is through deep learning, a machine learning technique that attempts to model high level abstractions. Algorithms like convolutional and artificial neural networks (CNNs, ANNs) have been used for galaxy morphology classification with impressive accuracy \citep{Ball2004, 
	Banerji2010, 
	Dieleman2015, 
	HuertasCompany2015}. 
A drawback to all machine learning classification techniques is the need for 
standardized training data, with more complex algorithms requiring more data. Furthermore, these data must be consistent for each survey: differences in resolution and depth can be implicitly learned by the algorithm making their application to disparate surveys challenging.  

In this work we present a system that preserves the best features of both visual and automatic classifications, developing for the first time a framework that brings both human and machine intelligence to the task of galaxy morphology to handle the scale and scope of next generation data. We demonstrate the effectiveness of such a system through a re-analysis of visual galaxy morphology classifications collected during the Galaxy Zoo 2 project, and combine these with a Random Forest machine learning algorithm that trains on a suite of non-parametric morphology indicators widely used for automated morphologies. The primary goal of this paper is to generalize how such a system would work in the context of upcoming surveys like LSST and Euclid.
As a proof of concept, we focus on the first question of the Galaxy Zoo decision tree. We demonstrate that our current implementation provides at least a factor of 8 increase in the rate of galaxy morphology classification while maintaining at least 93.5\% classification accuracy as compared to Galaxy Zoo 2 published data. We first present an overview of our framework, which also serves as a blueprint for this paper.

\added{\subsection{Galaxy Zoo Express Overview}}

The Galaxy Zoo Express (GZX) framework combines human and machine to  increase morphological classification efficiency, both in terms of the classification rate and required human effort. Figure~\ref{fig: schematic} presents a schematic of GZX including section numbers as a shortcut for the reader. We note that transparent portions of the schematic represent areas of future work which we explore in Section~\ref{sec: visions}. Any system combining human and machine classifications will have a set of generic features: a group of human classifiers, at least one machine classifier, and a decision engine which determines how these classifications should be combined.

In this work we demonstrate our system through a re-analysis of Galaxy Zoo 2 (GZ2) crowd-sourced classifications \added{as described in Section \ref{sec: data}. We compute ``ground truth'' labels for each galaxy in the GZ2 sample from the published GZ2 classification catalogue (Section \ref{sec: ground truth}).} The GZ2 data allow us to create simulations of human classifiers \deleted{(described in Section~\ref{sec: data})} whose classifications are used most effectively when processed with SWAP, a Bayesian code first developed for the Space Warps gravitational lens discovery project~\citep{Marshall2016} and described in Section~\ref{sec: SWAP}. \replaced{Galaxy images classified through SWAP (hereafter, \textit{subjects}) provide the machine's training sample.}{SWAP aggregates the crowd-sourced classifications of galaxy images (hereafter, \textit{subjects}) producing a final label for each subject (Section \ref{sec: fiducial}). \added{We show that SWAP produces significant gains in classification efficiency as well as a reduction of human effort in Sections \ref{sec: swap is faster} and \ref{sec: less human effort}.} In Section \ref{sec: swap gz2 disagree} we compare these labels to the ``ground truth'' labels computed from GZ2's traditional crowd-sourced classification method. Subjects classified by SWAP then provide the machine's training sample.}

In Section~\ref{sec: machine}, we incorporate a machine classifier. We develop a Random Forest algorithm that trains on measured morphology indicators such as Concentration, Asymmetry, Gini coefficient and \M{20}, well-suited for the top-level question of the GZ2 decision tree, discussed below. \added{Section \ref{sec: decision engine} discusses the decision engine we develop that delegates tasks between human classification and the Random Forest.} After a sufficient number of subjects have been classified by humans \added{via SWAP}, the machine is trained and its performance assessed through cross-validation. This procedure is repeated nightly and the machine's performance increases with the size of the training sample, albeit with a performance limit. Once the machine reaches an acceptable level of performance it is applied to the remaining galaxy sample \added{as explored in Section \ref{sec: machine shop}}. 

\added{The results of our combined GZX system are provided in Section \ref{sec: results}.} Even with this simple description, one can see that the classification process will progress in three phases. First, the machine will not yet have reached an acceptable level of performance; only humans contribute to subject classification. Second, the machine's performance will improve; both humans and machine will be responsible for classification. Finally, machine performance will slow; remaining images will likely need to be classified by humans. This result is detailed in Section \ref{sec: who retires what}. Furthermore, in Section \ref{sec: machine performance}, we find evidence that the Random Forest may be capable of correctly identifying subjects that humans miss providing a complimentary approach to galaxy classification. \deleted{These results are explored in  Section~\ref{sec: results}.} This blueprint allows even modest machine learning routines to make significant contributions alongside human classifiers and removes the need for ever-increasing performance in machine classification. \added{Discussion and conclusions are presented in Section \ref{sec: visions}.}

\section{Galaxy Zoo 2 Classification Data} 
\label{sec: data}

Our simulations utilize original classifications made by volunteers during the GZ2 project. These data\footnote{\url{data.galaxyzoo.org}} are described in detail in~\cite{Willett2013}, though we provide a brief overview here.  The GZ2 subject sample consists of 285,962 galaxies identified as the brightest 25\% ($r$-band magnitude $< 17$) residing in the SDSS North Galactic Cap region from Data Release 7 and included subjects with both spectroscopic and photometric redshifts out to $z < 0.25$. Subjects were shown as colour composite images via a web-based interface\footnote{\url{www.galaxyzoo.org}} wherein volunteers answered a series of questions pertaining to the morphology of the subject. With the exception of the first question, subsequent queries were dependent on volunteer responses from the previous task creating a complex decision tree\footnote{A visualization of this decision tree can be found at \url{https://data.galaxyzoo.org/gz_trees/gz_trees.html}}. Using GZ2 nomenclature, a \textit{classification} is the total amount of information about a subject obtained by completing all tasks in the decision tree. A subject is \textit{retired} after it has achieved a sufficient number of classifications.

For our current analysis, we choose the first task in the tree: ``Is the galaxy simply smooth and rounded, with no sign of a disk?" to which possible responses include ``smooth", ``features or disk", or ``star or artifact". This choice serves two purposes: 1) this is one of only two questions in the GZ2 decision tree that is asked about every subject thus maximizing the amount of data we have to work with, and 2) our analysis assumes a binary task and this question is simple enough to cast as such. Specifically, we combine ``star or artifact" responses with ``features or disk" responses.

\added{\subsection{``Ground truth'' labels}}
\label{sec: ground truth}

We assign each subject a descriptive label in order to validate our classification output against that of GZ2. GZ2 classifications are composed of volunteer vote fractions for each response to every task in the decision tree, denoted as $f_{\mathrm{response}}$.  \replaced{They are derived from the fraction of volunteers who voted for a particular response and are thus approximately continuous.}{The most basic of these is computed simply as $f_{\mathrm{r}} = n_{\mathrm{r}}/n_{\mathrm{t}}$, that is, the number of votes of response $r$ divided by the total number of votes for task $t$. Vote fractions are thus approximately continuous.} A common technique is to place a threshold on these vote fractions to select samples with an emphasis on purity or completeness, depending on the science case. For our current analysis we choose a threshold of 0.5, that is, if \ffeat+\fstar~$ >$ \fsmooth, the galaxy is labelled~\feat, otherwise it is labelled~\notfeat. We note that only 512 subjects in the GZ2 catalogue have a majority \fstar, contributing less than half a percent contamination when combining the ``star or artifact" with ``features or disk" responses.

The GZ2 catalogue publishes three types of vote fractions for each subject: raw, weighted, and debiased. Debiased vote fractions are calculated to correct for redshift bias, a task that GZX does not perform. The weighted vote fractions account for inconsistent volunteers. The SWAP algorithm (described below) also has a mechanism to weight volunteer votes, however, the two methods are in stark contrast. For consistency, we thus derive labels from the \replaced{raw vote fractions (\raw); those that have received no post-processing whatsoever.}{simple ``raw'' vote fractions defined above, and designate the resulting labels as \raw.} In total, the data consist of over 14 million classifications from 83,943 individual volunteers. 

The \added{\raw} labels we compute from GZ2 vote fractions are used solely to validate our classification method and are thus considered ``ground truth,'' though this is, of course, subjective. Furthermore, we envision our framework being applied to never-before-classified image sets for which ``ground truth" labels would not yet exist. Nevertheless, in Appendix~\ref{sec: vary threshold} we show how different choices of our descriptive GZ2 labels change the perceived quality of our classification system and demonstrate that our method yields robust galaxy classifications.

\section{Efficiency through intelligent human-vote aggregation}
\label{sec: SWAP}

Galaxy Zoo 2 did not have a predictive retirement rule, rather each galaxy received a median of 44 independent classifications. Once the project reached completion, inconsistent volunteers were down-weighted~\citep{Willett2013}, a process that does not make efficient use of those who are exceptionally skilled. To intelligently manage subject retirement and increase classification efficiency, we adapt an algorithm from the Zooniverse project Space Warps~\citep{Marshall2016}, which searched for and discovered several gravitational lens candidates in the CFHT Legacy Survey~\citep{More2016}.  Dubbed SWAP (Space Warps Analysis Pipeline), this algorithm computed the probability that an image contained a gravitational lens given volunteers' classifications and experience after being shown a training sample consisting of simulated lensing events.  We provide \replaced{a brief}{an} overview here; \added{ interested readers are encouraged to refer to \cite{Marshall2016} for additional details}. 

\added{\subsection{The SWAP algorithm}}

\added{SWAP evaluates the accuracy of individual classifiers based on their responses to subjects where the true classification is known, and applies those evaluations to the consensus classifications of subjects where the true classification is unknown in order to improve classification efficiency and reduce the classification effort required to complete a project. In order to achieve this,} SWAP assigns each volunteer an \textit{agent} which interprets that volunteer's classifications. Each agent assigns a 2$\times$2 confusion matrix to their volunteer which encodes that volunteer's probability to correctly identify feature \A~given that the subject exhibits feature \A; and the probability to correctly identify the absence of feature \A~(denoted \N) given that the subject does not exhibit that feature. The agent updates these probabilities by estimating them as 

\begin{equation}
P(``X" | X, \mathbf{d}) \approx \frac{\mathcal{N}_{``X"}}{\mathcal{N}_{X}}
\end{equation}
where $X$ is the true classification of the subject and ``$X$" is the  classification made by the volunteer upon viewing the subject. Thus $\mathcal{N}_{``X"}$ is the number of classifications the volunteer labelled as type $X$, $\mathcal{N}_X$ is the number of subjects the volunteer has seen that were actually of type $X$, and $\mathbf{d}$ represents the history of the volunteer, i.e., all subjects they have seen. Therefore the confusion matrix for a single volunteer goes as

\begin{eqnarray}
\mathcal{M} & = & \left[
	\begin{array}{cc}
		P(``A"|N, \mathbf{d}) ~~& P(``A" | A, \mathbf{d}) \\[0.3em]
		P(``N"|N, \mathbf{d})~~& P(``N"|A, \mathbf{d}) \\[0.3em]
	\end{array}\right]
\end{eqnarray}
where probabilities are normalised such that $P(``A"|A) = 1- P(``N" | A) $.

Each subject is assigned a prior probability that it exhibits feature \A: $P(A) = p_0$. When a volunteer makes a classification, Bayes' theorem is used to compute how that subject's prior probability should be updated into a posterior using elements of the agent's confusion matrix. As the project progresses, each subject's posterior probability is updated after every volunteer classification, nudged higher or lower depending on volunteer input. Upper and lower probability thresholds can be set such that when a subject's posterior crosses the upper threshold it is highly likely to exhibit feature \A; while if it crosses the lower threshold it is highly likely that feature \A~is absent. Subjects whose posteriors cross either of these thresholds are considered retired.

\subsection{Gold-standard sample}\label{sec: training sample}

A key feature of the original Space Warps project was the training of 
individual volunteers through the use of simulated images. These were interspersed with real imaging and were predominantly shown at the beginning of a volunteer's engagement with the project, allowing that volunteer's agent time to update before classifying real data. Volunteers were provided feedback in the form of a pop-up comment after classifying a training image. GZ2 did not train volunteers in such a way, presenting a challenge when applying SWAP to GZ2 classifications. Though we cannot retroactively train GZ2 volunteers, we develop a gold standard sample and arrange the order of gold standard classifications in order to mimic the Space Warps system.

\begin{figure}[t!]
\includegraphics[width=3.45in]{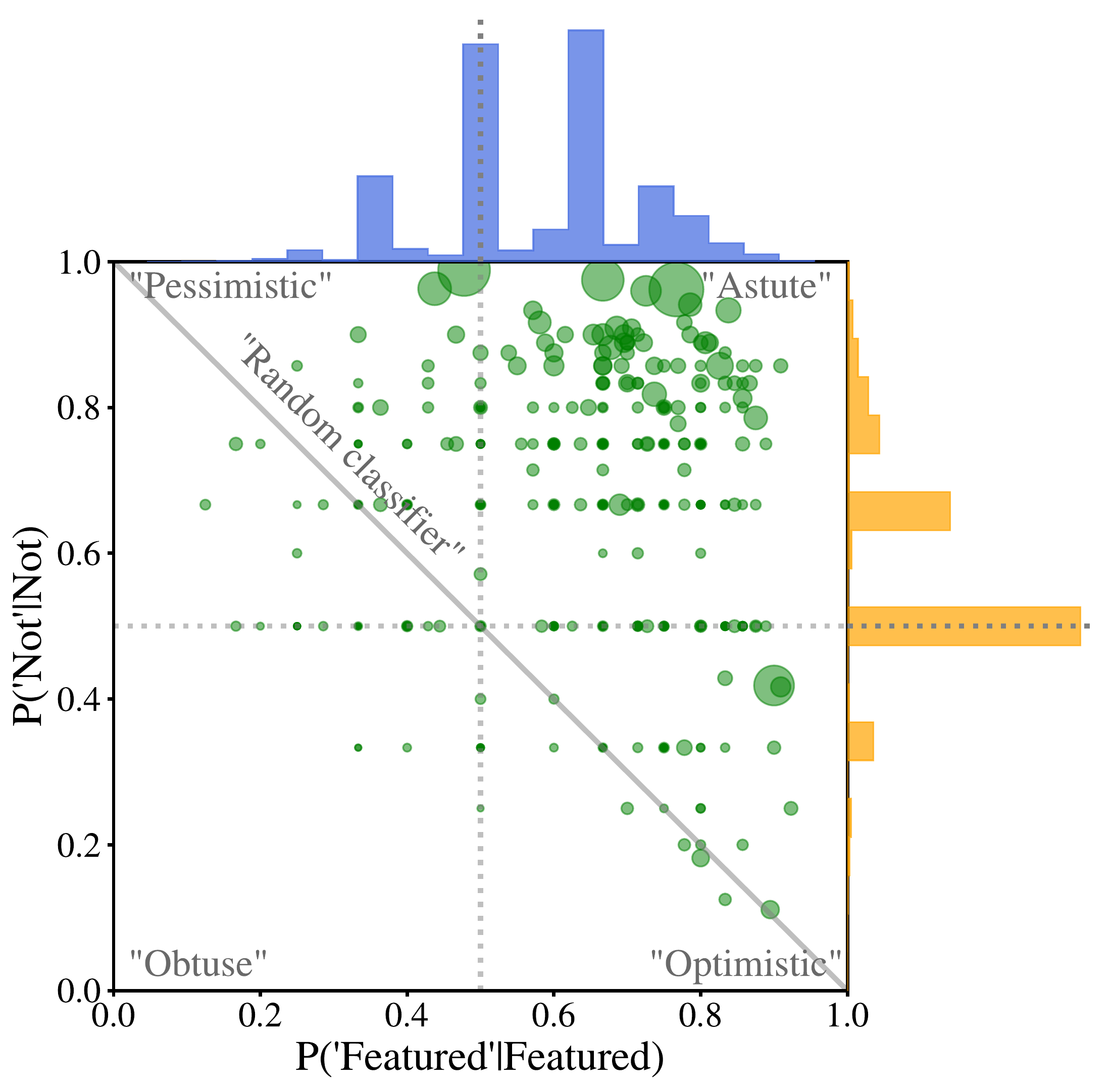}
\caption{Confusion matrices for 1000 randomly selected GZ2 volunteers after fiducial SWAP assessment. Circle size is proportional to the number of gold standard subjects each volunteer classified. The histograms on top and right represent the distribution of each component of the confusion matrix for all volunteers.  A quarter of GZ2 volunteers are ``Astute": they correctly identify both \feat~and \notfeat~subjects more than 50\% of the time. The peaks at 0.5 in both distributions are due primarily to volunteers who see only one training image: only half of their confusion matrix is updated. \label{fig: volunteer training}}
\end{figure}

\begin{figure}[t!] 
\centering
\includegraphics[width=3.25in]{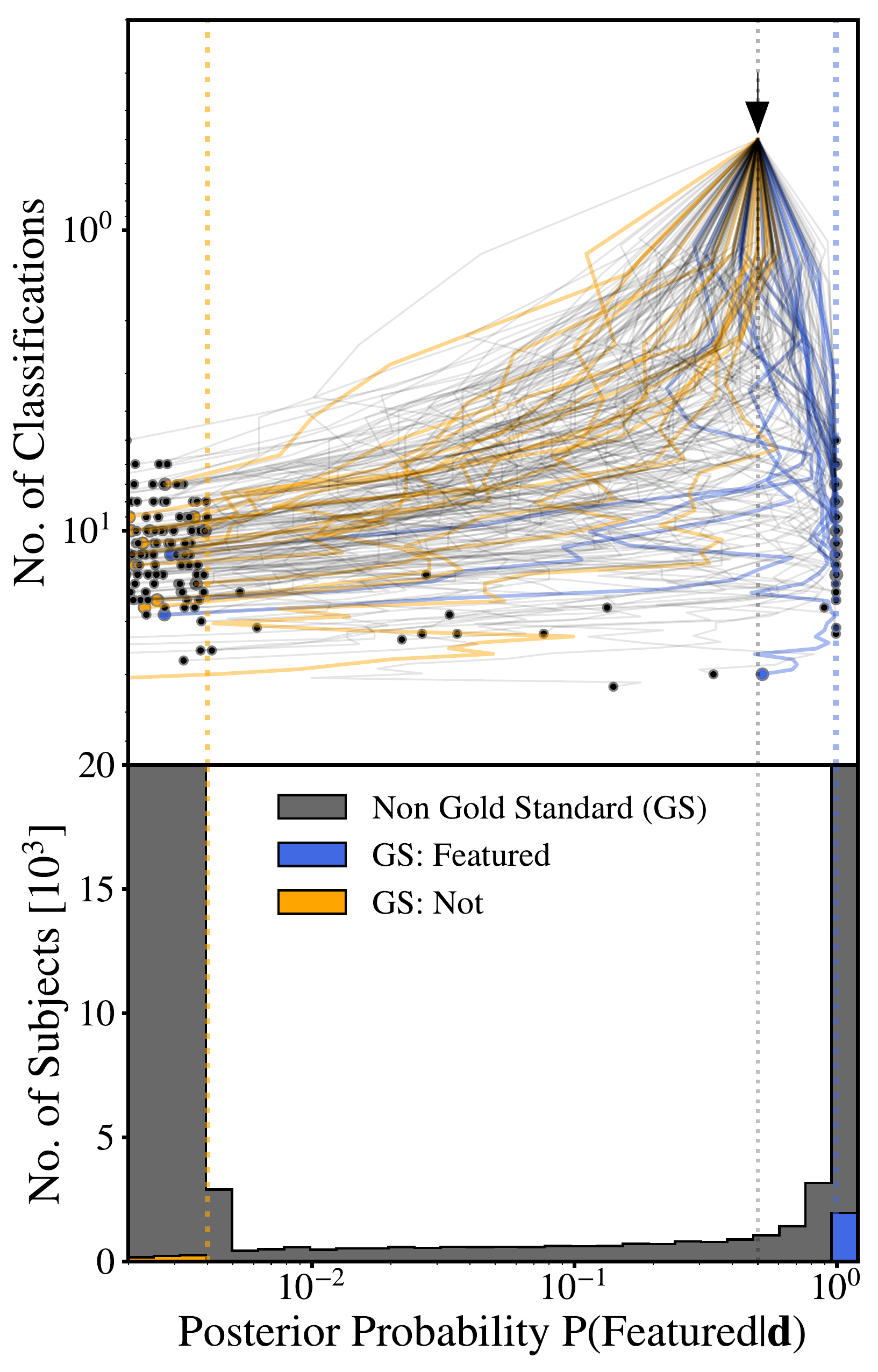}
\caption{Posterior probabilities for GZ2 subjects.  The top panel depicts the probability trajectories of 200 randomly selected GZ2 subjects. All subjects begin with a prior of 0.5 denoted by the arrow. Each subject's probability is nudged back and forth with each volunteer classification. From left to right the dotted vertical lines show the \notfeat~threshold, prior probability, and \feat~threshold. Different colours denote different types of subjects. The bottom panel shows the distribution in probability for all GZ2 subjects by the end of our simulation, where the y axis is truncated to show detail.  \label{fig: subject probabilities}}
\end{figure}

We create a gold standard sample by selecting 3496 SDSS galaxies representative of the relative abundance of T-Types, a numerical index of a galaxy's stage along the Hubble sequence, at $z\sim0$ by considering galaxies that overlap with the~\cite{NairAbraham2010} catalogue, a collection of $\sim$14K galaxies classified by eye into T-Types. We generate new expert labels for these galaxies that are consistent with the labels we defined for GZ2 classifications. These are provided by 15 professional astronomers, including members of the Galaxy Zoo science team, through the Zooniverse platform.\footnote{The Project Builder template facility can be found at \url{http://www.zooniverse.org/lab.}}  The question posed was identical to the original top-level GZ2 question and at least five experts classified each galaxy. Votes are aggregated and a simple majority provides an expert label for each subject. This ensures that our expert labels are defined in exactly the same manner as the labels we assign the rest of the GZ2 sample. Our final dataset consists of the GZ2 classifications made by those volunteers who classify at least one of these gold standard subjects. We thus retain for our simulation 12,686,170 classifications from 30,894 unique volunteers. When running SWAP, classifications of gold standard subjects are always processed first.

\subsection{Fiducial SWAP simulation}
\label{sec: fiducial}

Before we run a simulation, a number of SWAP parameters must be chosen:  the initial confusion matrix for each volunteer's agent, (\Pf, \Pn); the subject prior probability, \p; and the retirement thresholds, \tf~and \tn. For our fiducial  simulation we initialize all confusion matrices at (0.5, 0.5), and set the subject prior probability, \p~$= 0.5$. We set  the~\feat~threshold, \tf, i.e., the minimum probability for a subject to be retired as~\feat, to $0.99$. Similarly, we set the~\notfeat~threshold, \tn~$= 0.004$. In Appendix~\ref{sec: tweaking swap} we show that varying these parameters has only a small affect on the SWAP output. To simulate a live project, we run SWAP on a time step of $\Delta t = 1$ day, during which SWAP processes all volunteer classifications with timestamps within that range. This is performed for three months worth of GZ2 classification data. Hereafter, we refer to this as \textit{GZ2 project time} where $0$ marks the first day of the original GZ2 project.

Figure~\ref{fig: volunteer training} (adapted from Figure 4 of~\citealt{Marshall2016}) demonstrates the volunteer assessment we achieve \added{at the end of our simulation}, and shows confusion matrices for 1000 randomly selected volunteers. The circle size is proportional to the number of gold standard subjects each volunteer classified. \added{If we were to examine this figure immediately prior to the start of classifications, it would show all points as small circles stacked precisely at the center of the figure since each volunteer is initially assigned a confusion matrix of (0.5, 0.5). As the simulation progresses, each volunteer's green circle is updated in both location and size according to their assessment of gold standard subjects until arriving at the figure shown here.} The histograms represent the distribution of each component of the confusion matrix for all volunteers. Nearly 25\% of volunteers are considered ``Astute"  indicating they correctly identify both \feat~and \notfeat~subjects more than 50\% of the time. Furthermore, as long as a volunteer's confusion matrix is different from a random classifier, they provide useful information to the project. The spikes at $0.5$ in the histograms are due to volunteers who see only one gold standard subject (i.e.,~\feat), leaving their probability in the other (\notfeat) unchanged. Additionally, 4\% of volunteers have a confusion matrix of (0.5, 0.5) indicating these volunteers classified two gold standard subjects of the same type, one correctly and one incorrectly.

\begin{figure}[t!] 
\centering
\includegraphics[width=3.2in]{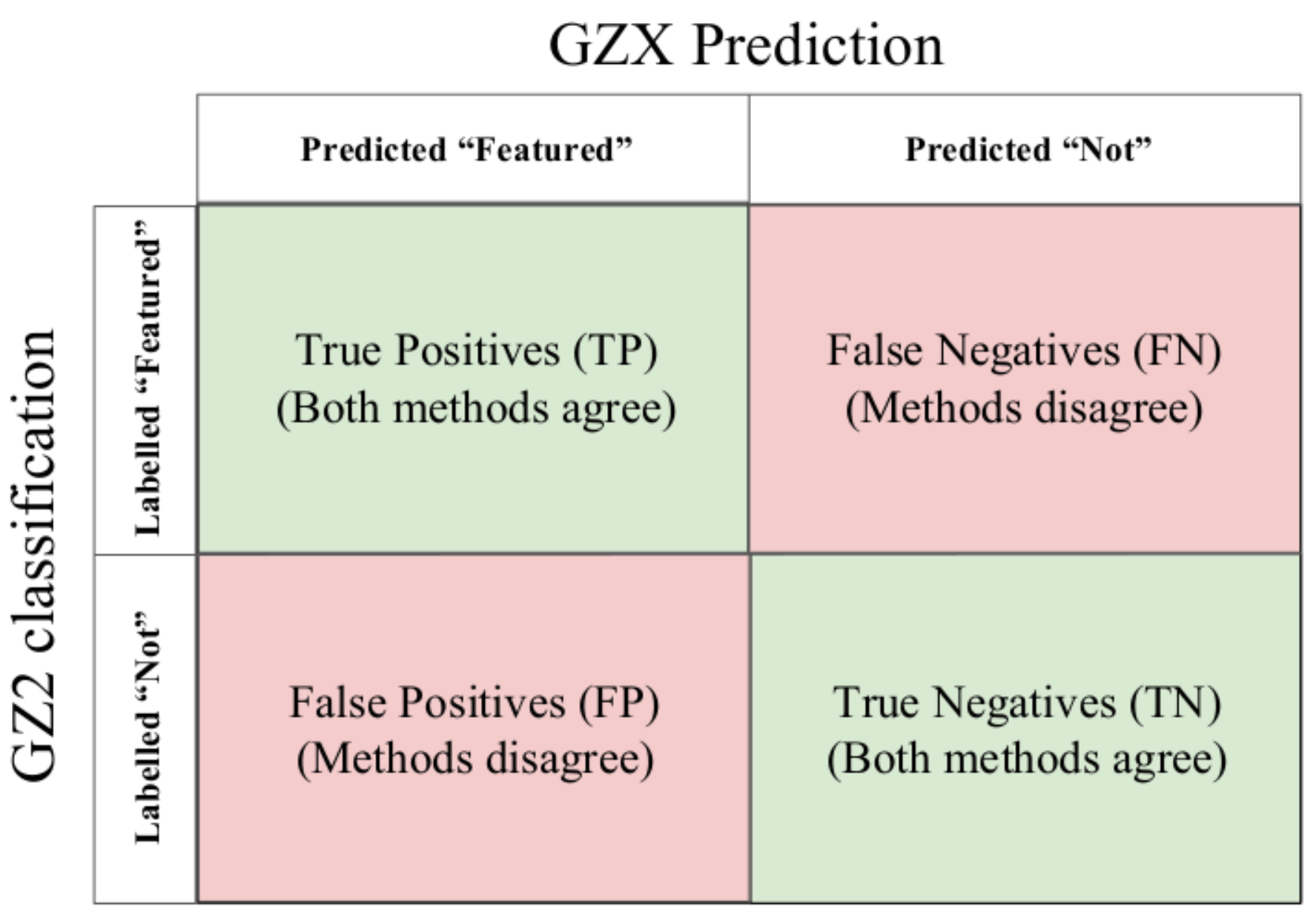}
\caption{Confusion matrix for comparing \replaced{GZ2 classifications to our method.}{our method to GZ2 which we consider to be ``ground truth'' as discussed in Section \ref{sec: ground truth}.}  True positives (TP) and true negatives (TN) indicate that the predictions from our method agree with GZ2 for subjects labelled \feat~and \notfeat, respectively. When the two classification methods disagree, the result is a sample of false negatives (FN) and false positives (FP). This allows us to easily compute  quality metrics like accuracy, completeness, and purity with respect to GZ2 as shown in Equations \ref{eqn: metrics}.} 
\label{fig: confusionmatrix}
\end{figure}

Figure~\ref{fig: subject probabilities} (adapted from Figure 5 of \citealt{Marshall2016}) demonstrates how subject posterior probabilities are updated with each classification. The arrow in the top panel denotes the prior probability, \p~$=0.5$. With each classification, that prior is updated into a posterior probability creating a trajectory through probability space for each subject. The blue and orange lines show the trajectories of a random sample of \feat~and \notfeat~subjects from our gold standard sample, while the black lines show the trajectories of a random sample of GZ2 subjects that were not part of the gold standard  sample. The blue and orange dashed lines correspond to the retirement thresholds, \tf~and \tn. The lower panel shows the full distribution of GZ2 subject posteriors at the end of our simulation, where the y-axis has been truncated to show detail. An overwhelming majority of subjects cross one of these retirement thresholds: of all subjects that SWAP ``sees", i.e., processes at least one classification, only 8\% have not reached retirement by the end of our simulation.

Our goal is to increase the efficiency of galaxy classification. We therefore  use as a metric the cumulative number of retired subjects as a function of GZ2 project time. We define a subject as GZ2-retired once it achieves at least 30 volunteer votes, encompassing 98.6\% of GZ2 subjects (this definition is quantified and its implications explored in Section~\ref{sec: swap is faster}).  In contrast, a subject is considered SWAP-retired once its posterior probability crosses either of the retirement thresholds defined above. 

However, it is important not to prioritize efficiency at the expense of quality. 
Because we have a binary classification, we can construct a confusion matrix from which we can compute the quality metrics of accuracy, completeness and purity as a function of GZ2 project time by comparing our predicted labels to the \raw~labels. Figure~\ref{fig: confusionmatrix} graphically ascribes semantic interpretations for the elements of this confusion matrix. From these we compute:

 \begin{align*}\label{eqn: metrics}
\mathrm{accuracy} &= \frac{TP + TN}{TP + FP + TN + FN} \\
\mathrm{completeness} &= \frac{TP}{TP +FN }\tag{3} \\
\mathrm{purity} &= \frac{TP}{TP + FP}
 \end{align*}

Thus a 100\% complete sample recovers \textit{all} subjects labelled \feat~by GZ2, whereas a 100\% pure sample recovers \textit{only} subjects labelled \feat~by GZ2. For example, by Day 20, SWAP retires 120K subjects with 96\% accuracy, 99.7\% completeness, and 92\% purity. 
 
\begin{figure}[t!]
\includegraphics[width=3.35in]{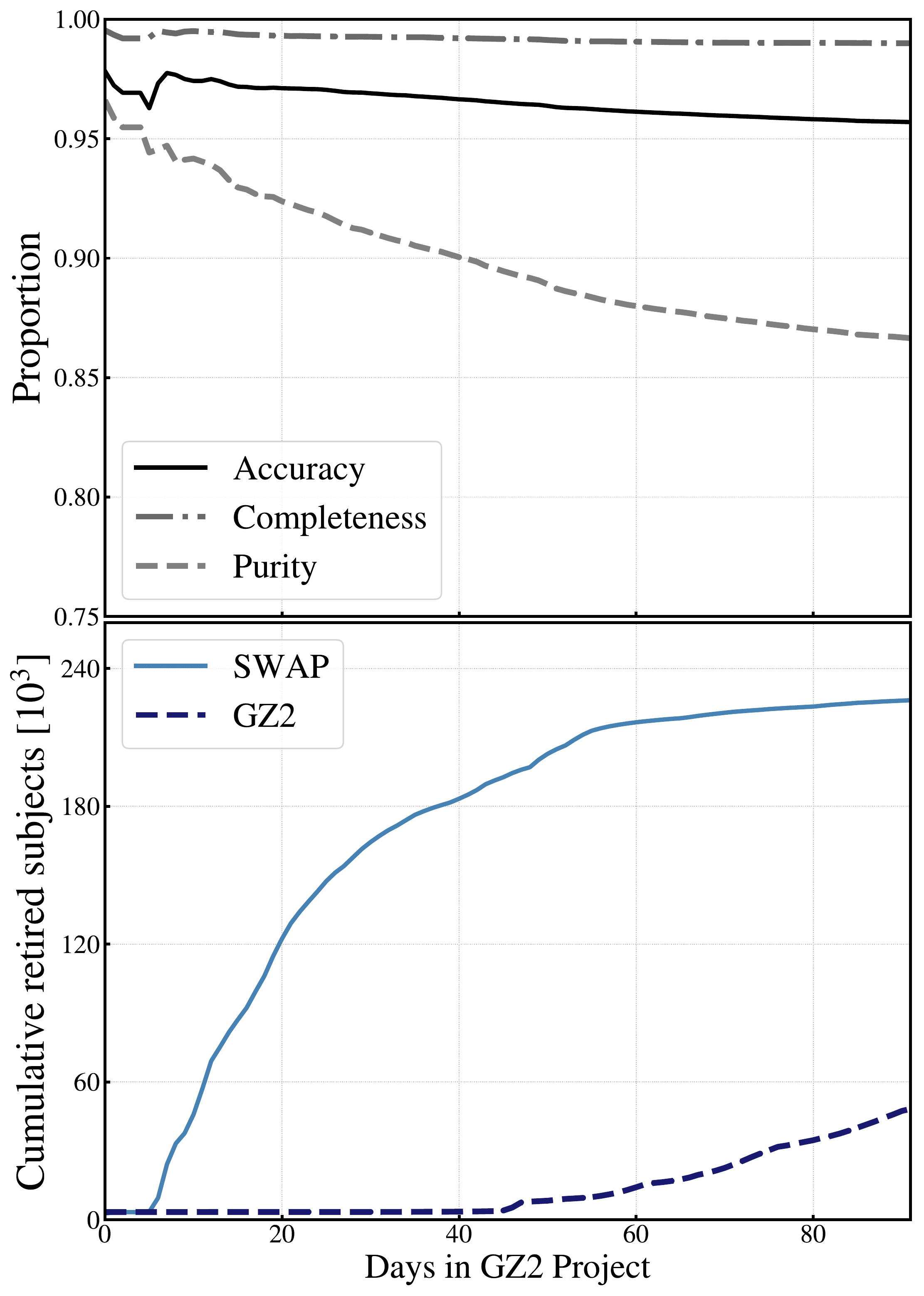}
\caption{Fiducial SWAP simulation demonstrates a factor of 4.7 increase in the rate of subject retirement as a function of GZ2 project time (bottom panel, light blue) compared with the original GZ2 project (dashed dark blue). After 92 days, SWAP retires over 226K subjects, while GZ2 retires $\sim$48K.  The top panel displays the quality metrics (greys). These are calculated by comparing labels predicted by SWAP to~\raw~labels (Section~\ref{sec: data}) for the subject sample retired by that day of the simulation. Thus, on the final day, SWAP retires 226,124 subjects with 95.7\% accuracy,  and with completeness and purity of~\feat~subjects at 99\% and 86.7\% respectively. The decrease in purity as a function of time is due, in part, to the fact that more difficult to classify subjects are retired later in the simulation (see Section~\ref{sec: swap is faster}). 
\label{fig: fiducial run}}
\end{figure}

Figure \ref{fig: fiducial run} and Table~\ref{tab: summary} detail the results of our fiducial SWAP simulation (``SWAP only") compared to the original GZ2 project. The bottom panel shows the cumulative number of retired subjects as a function of GZ2 project time. By the end of our simulation, GZ2 (dashed dark blue) retires $\sim$50K subjects while SWAP (solid light blue) retires 226,124 subjects. We thus classify 80\% of the entire GZ2 sample in three months. Processing volunteer classifications through SWAP presents nearly a factor of 5 increase in classification efficiency. The top panel of Figure~\ref{fig: fiducial run} demonstrates the quality of those classifications as a function of time and establishes that our full SWAP-retired sample is 95.7\% accurate, 99\% complete, and 86.7\% pure. We discuss these small discrepancies in Section~\ref{sec: swap gz2 disagree}.

\begin{figure*}[t!]
\includegraphics[width=7in]{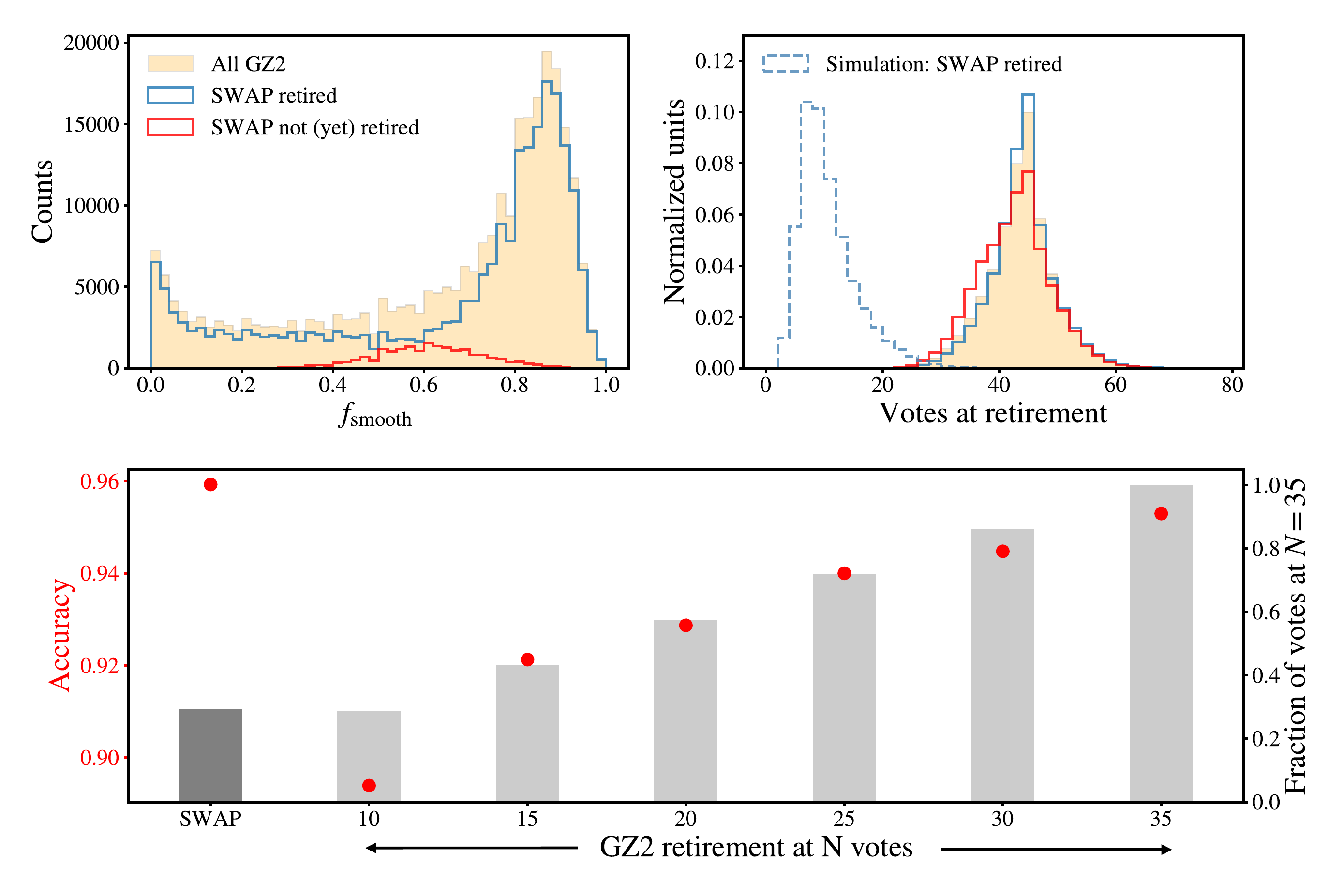}
\caption{SWAP's intelligent retirement mechanism requires only 30\% of the classifications that GZ2 needs for the top-level question due to SWAP's ability to retire easier subjects quickly, while more difficult subjects remain in the system to accrue additional classifications. 
\textit{Top panels:} The top left panel shows~\fsmooth~for the entire GZ2 sample (orange), the subjects retired by SWAP (blue), and subjects that SWAP has not yet retired by the end of our simulation (red). The latter distribution peaks at \fsmooth~$\sim 0.6$, which can intuitively be understood as the most difficult to classify subjects: those with \fsmooth~$\le 0.5$ are easily identified as \feat, while those with \fsmooth~$\ge 0.8$ are more obviously \notfeat. The top right panel provides additional evidence showing the number of votes at retirement for both the original GZ2 project (solid lines) and our SWAP simulation (dashed blue). The left-skew inherent in the red SWAP-not-yet-retired sample is due to difficult-to-classify subjects that received only 30-40 classifications during the GZ2 project. Even after processing all available classifications, SWAP cannot retire these subjects without additional volunteer input. 
\textit{Bottom panel:} Here we compare SWAP to results of simulations of GZ2 run with a lower retirement limit in order to evaluate whether or not GZ2's considerable number of votes per subject are necessary solely to populate subqueries. Solid bars show the number of classifications required to retire the same number of galaxies as SWAP (dark grey) for different fixed retirement limits in GZ2 (light grey). The height of the bars are normalised to show the counts relative to the highest simulated GZ2 retirement limit we test ($N=35$, right vertical axis).  The accuracy of the classifications for these simulated GZ2 runs against the full GZ2 project are shown as red points (left vertical axis). If GZ2 retirement were set at a level ($N=10$) that reproduces the total number of classifications logged by SWAP, the accuracy would be below 90\% (versus SWAP's 96\%).  Instead, GZ2 requires, at minimum, 3.5 times as many votes to approach the same accuracy (95\%) as SWAP. Simulated GZ2 sessions were run 100 times, randomly selecting subsamples with the same number of galaxies as were retired during our fiducial SWAP simulation. Quantities shown are averages of these trials; statistical error bars are too small to be seen.}
\label{fig: swap is faster}
\end{figure*}

\subsection{Intelligent subject retirement}
\label{sec: swap is faster}

That SWAP achieves a classification rate nearly 5 times faster than GZ2 comes with a caveat: we consider only the top-level question of the GZ2 decision tree. It can be argued that GZ2 did not need $\sim$40 votes per subject to achieve exquisite sampling for the top-level question but rather adequate sampling for the subqueries. It might therefore be the case that the top-level question could be accurately resolved with far fewer classifications. In order to put SWAP and GZ2 on equal footing we determine the minimum number of votes, $N$, that the GZ2 project would need in order to replicate the original GZ2 outcome for the top-level  classification task for a canonical 95\% of its sample.

We compute the raw vote fractions (\ffeat, \fsmooth, and \fstar) for every subject in the GZ2 sample using only the first $N$ classifications for $N \in [10, 15, 20, 25, 30, 35]$. From this, we compute descriptive labels as described in Section~\ref{sec: ground truth}. Our SWAP simulation did not retire every subject in the GZ2 sample. We therefore select 100 random subsamples each consisting of 226,124 subjects, and compute the accuracy and the total number of GZ2 classifications necessary to retire each subsample. These results are shown in the bottom panel of Figure~\ref{fig: swap is faster} for each value of $N$ along with the accuracy and total classifications for our SWAP simulation. We see that GZ2 needs at least 35 votes per subject in order to achieve consistent class labels 95\% of the time,  a full 3.5 times more classifications than  SWAP needs to achieve the same accuracy. Furthermore, this justifies our choice of defining a subject as GZ2-retired once it reaches at least 30 classifications.

SWAP's performance can be explained through its retirement mechanism. GZ2 did not have a predictive retirement rule, rather the project was declared complete when the median classification count for the ensemble reached a value that was deemed to be sufficient for accurate characterization of the classification. In contrast, SWAP retires ``easier" subjects first while harder subjects remain in the system for longer (requiring many more votes to nudge that subject's posterior across a retirement threshold).  Evidence for this can be seen in the top two panels of Figure~\ref{fig: swap is faster}. The top left panel shows the distribution of \fsmooth~for the entire GZ2 sample (orange), the SWAP-retired sample (blue), and the sample of subjects which SWAP has not yet retired, of which there are $\sim$19K at the end of our simulation. The SWAP-retired sample generally follows the same distribution as GZ2-full except for the noticeable dip around \fsmooth$=0.6$. In contrast, the SWAP-not-yet-retired sample peaks at \fsmooth$=0.6$. These subjects can be interpreted as being the most difficult to classify which can be understood intuitively: galaxies with \fsmooth~$\le 0.5$ are easily identified as having features, while galaxies with \fsmooth~$\ge 0.8$ are more obviously elliptical.

This is further corroborated in the top right panel of Figure \ref{fig: swap is faster} which shows the distribution of the number of classifications a subject had at the time of retirement. The solid lines show this distribution from the original GZ2 project for the same subsamples as the top left panel. For comparison, the dashed line shows the number of classifications at retirement realized during our SWAP simulation. Again, we see that the SWAP-retired sample is representative of GZ2 as a whole. However, the distribution for the SWAP-not-yet-retired sample is skewed toward fewer total classifications.

\begin{figure}[t!] 
\centering
\includegraphics[width=3.5in]{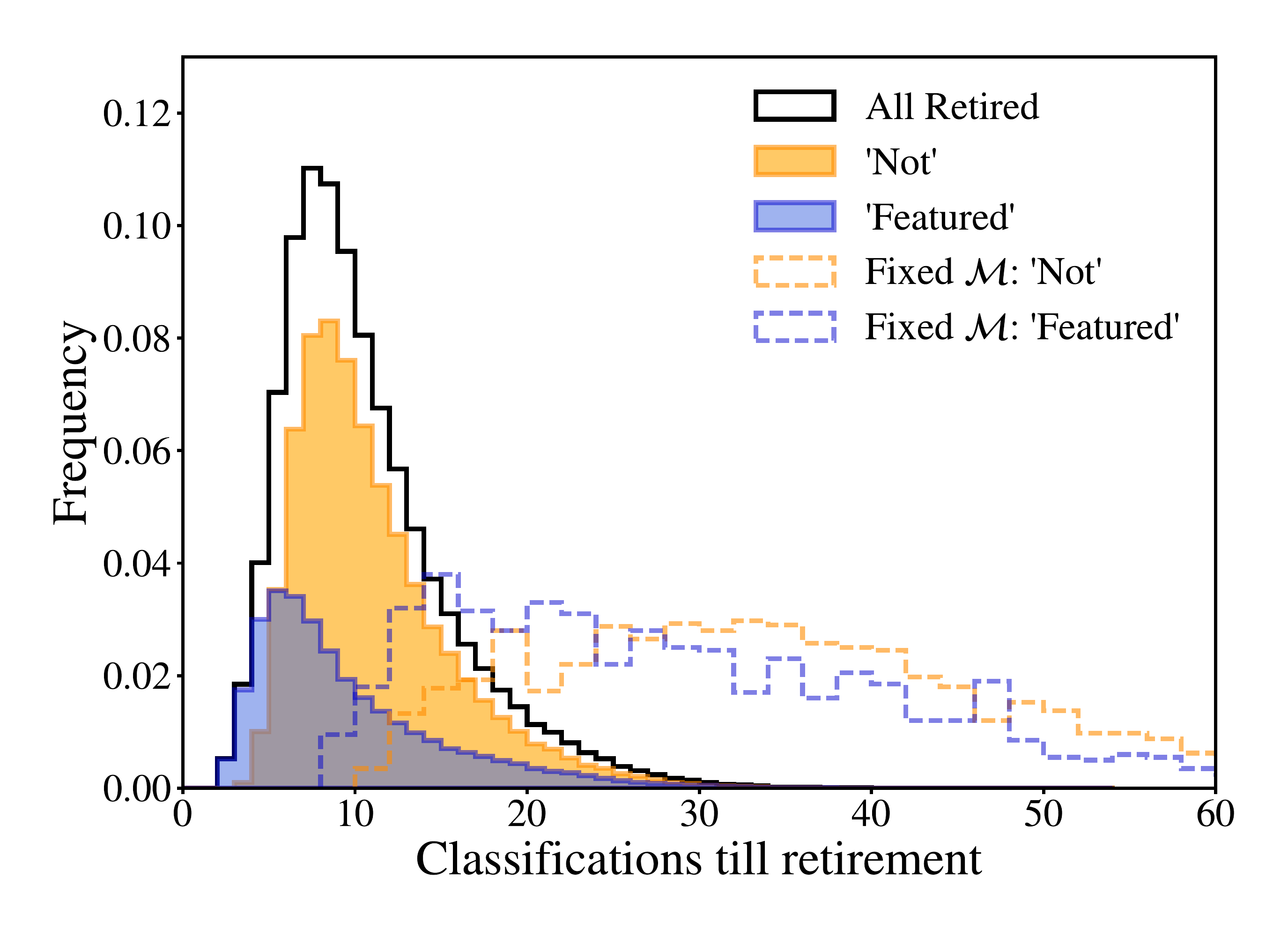}
\caption{
SWAP's volunteer-weighting mechanism provides a factor of three reduction in the human effort required to retire GZ2 subjects. The filled histograms show the number of volunteer classifications per subject achieved during our SWAP simulation broken down by class label, where the solid black line is the total. The dashed histograms are results from our toy model in which we simulate volunteers with fixed confusion matrices, effectively disengaging SWAP's volunteer-weighting mechanism. These broad distributions require $\sim$3 times more classifications per subject to reach the same retirement thresholds. } \label{fig: swap vote distributions}
\end{figure}

To understand this, consider the following: GZ2 served subject images at random with the  exception that, towards the end of the project, subjects with low numbers of classifications were shown at a higher rate \citep{Willett2013}. The median number of classifications was 44 with the full distribution shown in orange in the top right panel of Figure~\ref{fig: swap is faster}. Our SWAP simulation processes these classifications in the same order as the original project (with the exception that gold-standard subject classifications are processed first as described in Section~\ref{sec: training sample}). Because our simulations cycle through only 92 days of GZ2 data, there are three general scenarios for why a subject has not yet been retired through SWAP: 1) SWAP has seen only a few of the many classifications for a given subject and it is not yet enough to retire it, 2) SWAP has seen many of the classifications for a subject but that subject is difficult; if we ran the simulation longer to process the remaining GZ2 classifications, SWAP would eventually retire it, and 3) SWAP has seen most or all of the classifications for a subject but it is difficult and there are few or no remaining GZ2 classifications; without additional volunteer input, these subjects will never be retired by SWAP.

It is this third category that skews the red distribution towards fewer GZ2 votes. These are difficult-to-classify subjects that have only 30 - 40 GZ2 classifications, all of which are processed by SWAP, but these subjects remain unretired. This is an indication that such subjects should have continued to accrue classifications in order to reach strong consensus.

We have demonstrated that SWAP retires subjects intelligently: quickly retiring easy-to-classify subjects while allowing those that are more difficult to collect additional classifications. SWAP thus requires only 30\% of the votes that GZ2 needs and retires nearly 5 times as many subjects during the three months of GZ2 project time that we include in our simulation.

\subsection{Reducing human effort}
\label{sec: less human effort}

SWAP's intelligent retirement mechanism is characterised, in large part, by the way SWAP estimates volunteer classification ability. This in turn allows for a dramatic reduction in the amount of human effort (votes) required. To see this more clearly, we consider a toy model wherein we simulate volunteers with fixed confusion matrices. We simulate 1000 \feat~subjects and 1000 \notfeat~subjects each with prior, \p~$ = 0.5$. We simulate 100 volunteer agents all with the same fixed confusion matrix of (0.63, 0.65), where these values are computed as the average \Pf~and \Pn~from our assessment of real volunteers, excluding the spikes at 0.5. We generate volunteer classifications based on this confusion matrix (i.e., volunteers will correctly identify \feat~subjects 63\% of the time) and update the subject's posterior probability with each classification. We track how many classifications are required for each subject's posterior to cross either the \feat~or \notfeat~retirement thresholds.

The results are presented in Figure~\ref{fig: swap vote distributions}. The filled blue and orange histograms show the number of classifications per subject achieved from our SWAP simulation, where volunteer agent confusion matrices are those from Figure~\ref{fig: volunteer training}. The dashed blue and orange distributions are the results from our toy model. When SWAP accounts for volunteer ability, most subjects are retired with between 6 and 15 votes, with a median of 9 votes. In contrast, when every volunteer is given equal weighting, subjects require 16 to 45 votes with a median of 30 votes before crossing one of the retirement thresholds. Thus the volunteer weighting scheme embedded in SWAP can reduce the amount of human effort required to retire subjects by a factor of three.

This reduction will be, in part, a function of the number of gold standard subjects each volunteer sees.  Our gold standard sample was chosen to be representative of morphology rather than evenly distributed among GZ2 volunteers. We thus find that half of our volunteers classify only one or two gold standard subjects. That we achieve a factor of three reduction when only half of our volunteer pool has seen $\ge 2$ gold standard subjects suggests that an additional reduction of  human effort is possible with more extensive volunteer training.

\begin{figure}[t!]
\includegraphics[width=3.5in]{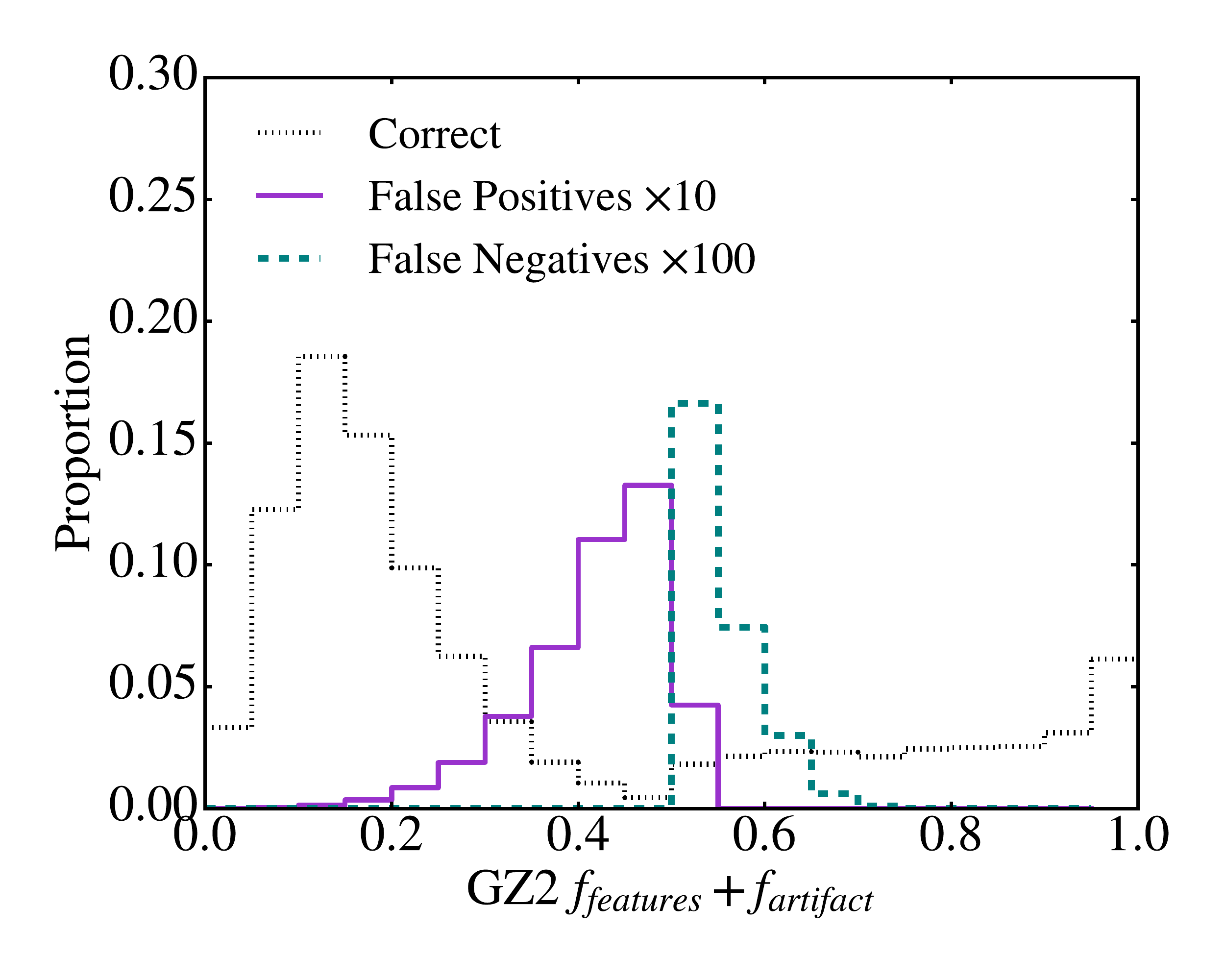}
\caption{Distribution of GZ2 \ffeat+\fstar~vote fractions for subjects correctly identified by SWAP (dotted grey), along with those identified as false positives (solid purple), and false negatives (dashed teal). The false positives and false negatives are scaled by factors of 10 and 100 respectively for easier comparison. From Section~\ref{sec: data}, subjects with values $> 0.5$ are defined as~\feat, however, the teal distribution indicates that SWAP labels them as~\notfeat. This is not necessarily a flaw of SWAP: 68.9\% of incorrectly identified subjects have $0.4 \le $~\ffeat +\fstar~$ \le 0.6$, nearly the same range as a 68\% confidence interval around our choosen threshold.  The overlap between the false positives and negatives is due to subjects that are exactly 50-50; by default these are labelled~\notfeat. \label{fig: SWAP sucks}}
\end{figure}

\subsection{Disagreements between SWAP and GZ2}
\label{sec: swap gz2 disagree}

Galaxy Zoo's strength comes from the consensus of dozens of volunteers voting on each subject. Processing votes with SWAP reduces the number of classifications to reach consensus. Though we typically recover the~\raw~label, SWAP disagrees about 5\% of the time. We thus examine the false positives (subjects SWAP labels as~\feat~but~\raw~labels as~\notfeat) and false negatives (subjects SWAP labels as~\notfeat~but~\raw~labels as~\feat). We explore these subjects in redshift, magnitude, physical size, and concentration but find no correlation with any of these variables, suggesting that, at least for this galaxy sample, the reliability of morphology depends on factors that are not captured by these coarse measurements. This is perhaps unsurprising since GZ2 subjects were selected from the larger GZ1 sample to be the brightest, largest and nearest galaxies:  precisely those subjects most accessible for visual classification. 

Instead we consider the stochastic nature of GZ2 vote fractions, which can be estimated as binomial. Let success be a response of ``smooth'' and failure be any other response. The $68\%$ confidence interval on a subject with \fsmooth~$=0.5$ is then $(0.42, 0.57)$ assuming 40 classifications, each with a probability of 0.5. Figure~\ref{fig: SWAP sucks} shows the distribution of~\ffeat+\fstar~for the false positives (solid purple), and the false negatives (dashed teal) compared to the  subjects where SWAP and GZ2 agree (dotted grey).  Recall that if this value is greater than 0.5, the subject is labelled~\feat. The majority of disagreements between SWAP and GZ2 are for subjects that have $0.4 <$~\ffeat+\fstar~$< 0.6$. It is thus unsurprising that SWAP and GZ2 disagree most within the approximate confidence interval of our selected GZ2 threshold. We note that the distribution overlap between false positives and false negatives is due to subjects that do not have a majority; these are labelled~\notfeat~by default. 

Two other effects contribute to the disagreement between SWAP and GZ2. First, as the number of classifications used to retire a galaxy decreases, the likelihood of misclassification by random chance increases. Second, disagreement arises due to expert-level volunteers whose confusion matrices are close to 1.0. These volunteers are essentially more strongly weighted, allowing that subject's posterior to cross a retirement threshold in as few as two classifications. In rare cases, despite training, some expert-level 
volunteers get it wrong compared to the gold-standard labels. These issues can be mitigated by requiring each subject reach a minimum number of classifications in addition to its posterior probability crossing a retirement threshold, thus combining the best qualities of GZ2 and SWAP.

\subsection{Summary}
We demonstrate nearly a factor of five increase in the classification rate, a reduction of at least a factor of three in the human effort necessary to maintain that increased rate, all while maintaining 95\% accuracy, nearly perfect completeness of~\feat~subjects, and with a purity that can be controlled by careful selection of input parameters to be better than 90\% (see Appendix~\ref{sec: tweaking swap}). Exploring those subjects wherein SWAP and GZ2 disagree, we conclude that the majority of this disagreement stems from the stochastic nature of~\raw~labels. We now turn our focus towards incorporating a machine classifier utilizing these SWAP-retired subjects as a training sample.

\section{Efficiency through incorporation of machine classifiers} \label{sec: machine}

We construct the full Galaxy Zoo Express by incorporating supervised learning, the machine learning task of inference from labelled training data. The training data consist of a set of training examples, and must include an input feature vector and a desired output label.  Generally speaking, a supervised learning algorithm analyses the training data and produces a function that can be mapped to new examples. A properly optimized algorithm will correctly determine class labels for unseen data. By processing human classifications through SWAP, we obtain a set of binary labels by which we can train a machine classifier. We briefly outline the technical details of our machine below,  turning towards the decision engine we develop in Section~\ref{sec: decision engine}. 

\subsection{Random Forests}

We use a Random Forest (RF) algorithm~\citep{Breiman2001},  an ensemble classifier that operates by bootstrapping the training data and constructing a multitude of individual decision tree algorithms, one for each subsample.  An individual decision tree works by deciding which of the input features best separates the classes. It does this by performing splits on the values of the input feature that minimize the classification error. These feature splits proceed recursively. Decision trees alone are prone to over-fitting, precluding them from generalising well to new data. Random Forests mitigate this effect by combining the output labels from a multitude of decision trees.  Specifically, we use the \texttt{RandomForestClassifier} from the Python module \texttt{scikit-learn} \citep{scikit-learn}.

\subsection{Grid Search and Cross-validation}

Of fundamental importance is the task of choosing an algorithm's hyperparameters, values which determine how the machine learns.   For a RF, key quantities include the maximum depth of individual trees (\texttt{max\_depth}), the number of trees in the forest (\texttt{n\_estimators}), and the number of features to consider when looking for the best split (\texttt{max\_features}). The goal is to determine which values will optimize the machine's performance and thus these values cannot be chosen \textit{a priori}. We perform a grid search with $k$-fold cross-validation whereby the training sample is split into $k$ subsamples. One subsample is withheld to estimate the machine's performance while the remaining data are used to train the machine. This is performed $k$ times and the average performance value is recorded. The entire process is repeated for every combination of the hyperparameters in the grid space and values that optimize the output are chosen. In this work we let $k=10$, however, we leave this as an adjustable input parameter. In the interest of computational speed, we set \texttt{n\_estimators} $=30$ and perform the grid search for \texttt{max\_depth} over the range $[5,16]$, and \texttt{max\_features} over the range $[\sqrt{D}, D]$, where $D$ is the number of features in the feature vector, described below.

\subsection{Feature Representation and Pre-Processing}

The feature vector on which the machine learns is composed of $D$ individual numeric quantities associated with the subject that the machine uses to discern that subject from others in the training sample. To segregate~\feat~from~\notfeat, we draw on ZEST \citep{Scarlata2007} and compute concentration, asymmetry, Gini coefficient, and M$_{20}$, the second-order moment of light for the brightest 20\% of galaxy pixels, as measured from SDSS DR12 $i$-band imaging (see Appendix \ref{sec: measuring morphology}). Coupled with SExtractor's measurement of ellipticity \citep{sextractor}, we provide the machine with a $D=5$ dimensional morphology parameter space.  These non-parametric diagnostics have long been used to distinguish between early- and late-type galaxies in an automated fashion \cite[e.g.,][]{Abraham1996, Bershady2000, Conselice2000, Abraham2003, Conselice2003, Lotz2004, Snyder2015}. Because the RF algorithm handles a variety of input formats, the only pre-processing step we perform is the removal of poorly-measured morphological indicators, i.e. catastrophic failures.

\begin{figure}[t!]
\includegraphics[width=3.25in]{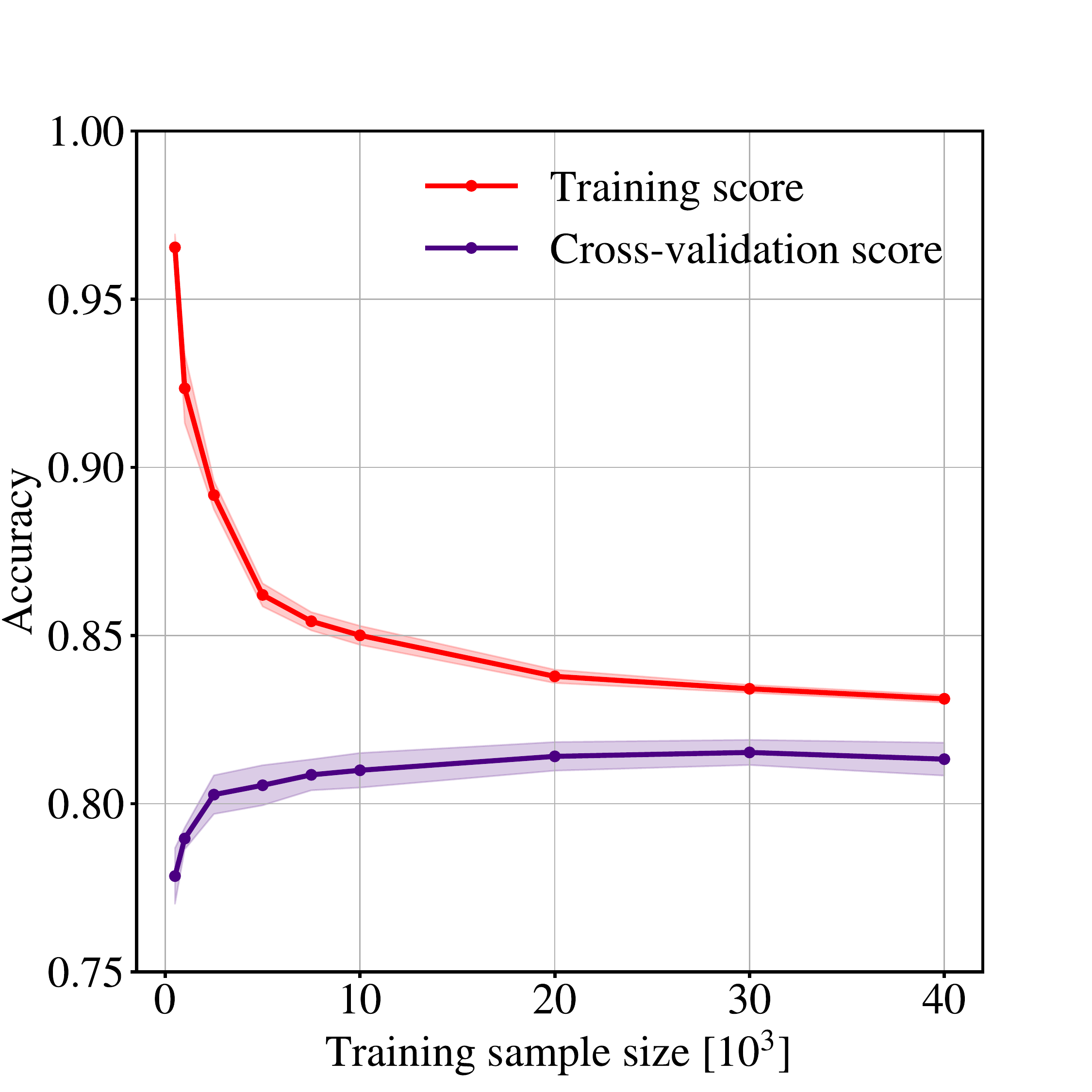}
\caption{Learning curve for a Random Forest with fixed hyperparameters. These curves show the mean accuracy computed during cross-validation and on the training sample, where the shaded regions denote the standard deviation. When the training sample size is small, the machine accurately identifies its own training sample but is unable to generalize to unseen data as evidenced by a low cross-validation score. This score increases with the size of the training sample but eventually plateaus indicating that larger training samples provide little in additional performance. \label{fig: learning curve}}
\end{figure}

\subsection{Decision Engine}
\label{sec: decision engine}

A number of decisions must be addressed before attempting to train the machine. 
In particular, which subjects should be designated as the training sample? 
When should the machine attempt its first training session? 
When has the machine's performance been optimized such that it will successfully
generalize to unseen subjects? The field of machine learning provides few hard rules 
for answering these questions, only guidelines and best practices. 
Here we briefly discuss our approach for the development of our decision engine.

As discussed in detail in Section~\ref{sec: SWAP}, SWAP yields a probability that a subject exhibits the feature of interest. While some machine algorithms can accept continuous input labels, the RF requires distinct classes. We thus use only those subjects which have crossed either of the retirement thresholds. Though we find that SWAP consistently retires 35-40\%~\feat~subjects on any given day of the simulation, a balanced ratio of~\feat~to~\notfeat~isn't guaranteed. Highly unbalanced training samples should be resampled to correct the imbalance; however, as we exhibit only a mild lopsidedness, we allow the machine to train on all SWAP-retired subjects.  

SWAP retires a few hundred subjects during the first days of the simulation.
In principle,  a machine can be trained with such a small sample, but will be unable
to generalize to unseen data. We estimate a minimum number of training samples
and the machine's ability to generalize by considering a learning curve, an illustration
of a machine's performance with increasing sample size for fixed hyperparameters. 
Figure~\ref{fig: learning curve} demonstrates such a curve wherein we plot
the accuracy from both the 10-fold cross-validation, and the trained machine
applied to its own training sample for a random sample of GZ2 subjects
required to be balanced between~\feat~and~\notfeat.  
We fix the RF's hyperparameters as follows: \texttt{max\_depth} $=8$, 
\texttt{n\_estimators } $=30$, and \texttt{max\_features} $=2$. 
When the sample size is small, the cross-validation score is low and the training 
score is high, a clear sign of over-fitting.  However, as the training 
sample size increases, the cross-validation score increases and eventually plateaus,
 indicating that larger training sets will yield little additional gain. 

We estimate this plateau begins when the training 
sample reaches 10,000 subjects and require SWAP retire at least this many 
 before the machine attempts its first training.  We estimate the machine 
has trained sufficiently if the cross-validation score fluctuates by less than 1\% 
for three consecutive nights of training to ensure we have reached the plateau.  
This requires that we record the machine's training performance each night, 
including how well it scores on the training sample, the 
cross-validation score, and the best hyperparameters.

\begin{table*}[]
	\centering
	\caption{Summary of key quantities for GZ2 and our various simulations. All quality metrics are calculated using~\raw~labels.}
	\label{tab: summary}
	\let\mc\multicolumn
	\begin{tabular}{lcccccc}
		
		\mc7c{ \textbf{Simulation Summary} } \\
		\hline \hline
			& Days	& Subjects Retired & Human Effort 	&  Accuracy 	& Purity 	& Completeness\\
		\mc2c{} 		& 	 	& (classifications) 	&  (\%)	    	& (\%)	& (\%)	\\
		\hline
			
		Galaxy Zoo 2	&	430 	& 285,962  	& 14,144,142 	& --   	& --    	 & --   \\
		SWAP only	&	92    	& 226,124          & 2,298,772	& 95.7 	& 86.7	 & 99.0     \\
		SWAP+RF   	& 32  	& 210,803 	& 936,887 	& 93.1    	& 83.2    	& 94.0      \\
		\hline
	\end{tabular}
\end{table*}

\begin{figure*}[t!]
\centering
\includegraphics[width=5.5in]{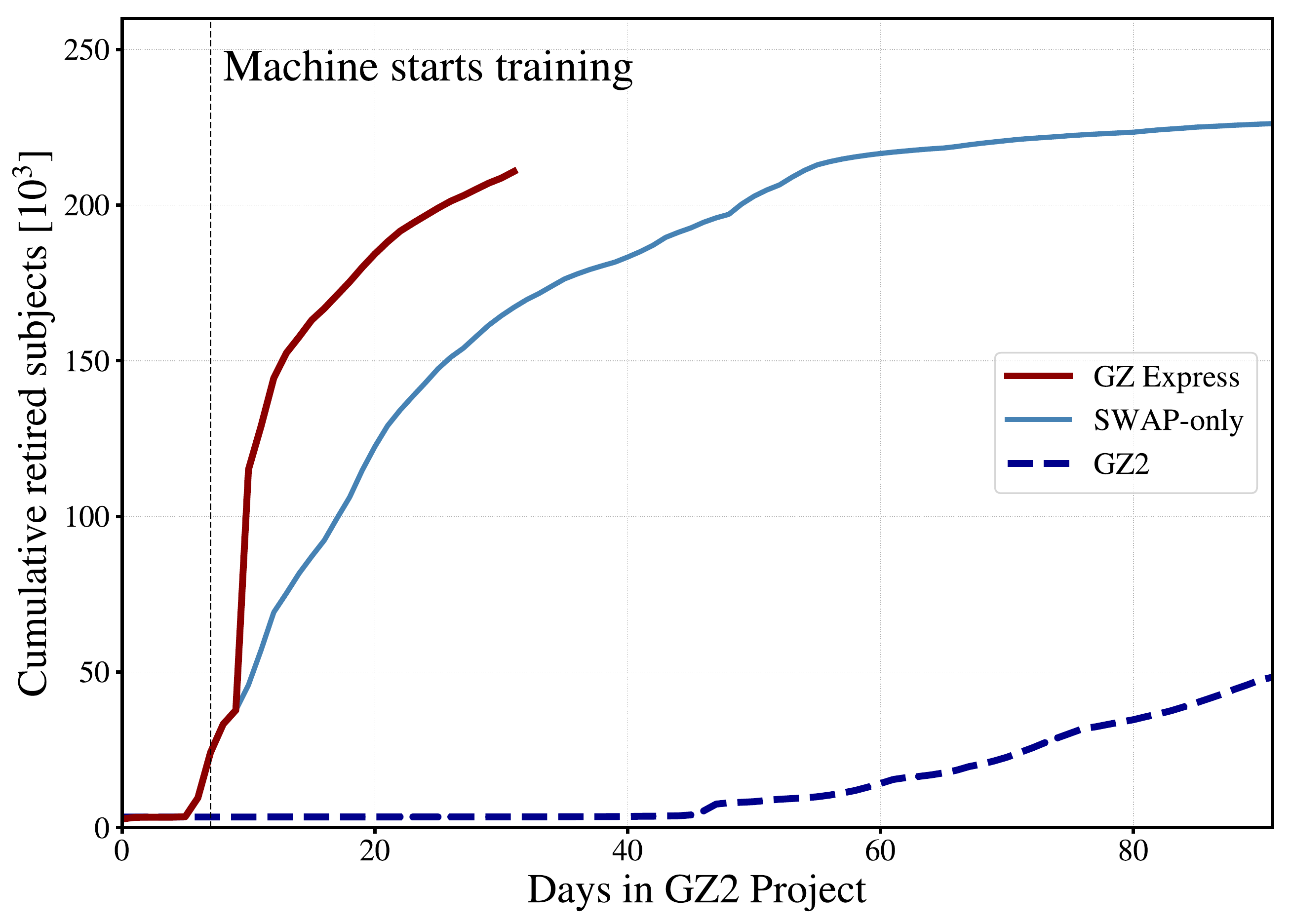}
\caption{By incorporating a machine classifier, GZX (red) increases the classification rate by an order of magnitude compared to GZ2 (dashed dark blue) and out-performs the SWAP-only run (light blue), retiring more than 200K subjects in just 27 days of GZ2 project time. The dashed black line marks the first night the machine trains. After several additional nights of training, it is deemed optimized and allowed to retire subjects. Both humans and machine then contribute to retirement. We end the simulation after 32 days having retired over 210K galaxies. See Table~\ref{tab: summary} for details. \label{fig: money}}
\end{figure*}

\subsection{The Machine Shop}
\label{sec: machine shop}

We can now describe a full GZX simulation, which begins with human classifications processed through SWAP for several days. Once at least 10K subjects have been retired, their feature vectors are passed to the machine for its inaugural training. A suite of performance metrics are recorded by a machine agent, similar in construction to SWAP's agents. This agent determines when the machine has trained sufficiently by assessing the variation in performance metrics for all previous nights of training. Once the machine has been optimized, the agent introduces it to the test sample consisting of any subject that has not yet reached retirement through SWAP and is not part of the gold standard sample.  

Analogous to SWAP, we generate a retirement rule for machine-classified subjects. In addition to the class prediction, the RF algorithm computes the probability for each subject to belong to each class.  This probability is simply the average of the probabilities of each individual decision tree, where the probability of a single tree is determined as the fraction of subjects of class X on a leaf node.  Only subjects that receive a class prediction of \feat~with $p_{\mathrm{machine}} \ge 0.9$ ($p_{\mathrm{machine}} \le 0.1$ for \notfeat) are considered retired. The remaining subjects have the possibility of being classified by humans or the machine on a future night of the simulation. This constitutes the core of our passive feedback mechanism. Subjects that are not retired by the machine can instead be retired by humans, thus providing the machine a more fully sampled morphology parameter space on future training sessions.

\section{Results} 
\label{sec: results}

We perform a full GZX simulation incorporating our RF with the fiducial SWAP run discussed in Section~\ref{sec: fiducial}. The machine attempts its first training on Day 8 with an initial training sample of $\sim$20K subjects. It undergoes several additional nights of training, each time with a larger training sample. By Day 12, SWAP has provided over 40K subjects for training and the machine's agent has deemed the machine optimized. The machine predicts class labels for the remaining 230K GZ2 subjects. Of those, the machine retires over 70K, dramatically increasing the subset of retired subjects. We end the simulation after 32 days, having retired $\sim$210K subjects as detailed in Table~\ref{tab: summary}. 

We present these results in Figure~\ref{fig: money} where subject retirement with GZX (red) is compared to our fiducial SWAP-only run (light blue) and GZ2 (dashed dark blue). Using the~\raw~labels as before, we compute our usual quality metrics on the full sample of GZX-retired subjects; reported in Table~\ref{tab: summary}. Accuracy and purity remain within a few percent of the SWAP-only run at \replaced{93.5\% and 84.2\%}{93.1\% and 83.2\%} respectively. Instead we see a 5\% decline in the completeness. While the SWAP-only run identified 99\% of~\feat~subjects, incorporation of the machine seems to miss a significant portion thus dropping GZX completeness to 94.0\%. We discuss this behaviour below.

By dynamically generating a training sample through a more sophisticated analysis of human classifications coupled with a machine classifier, we retire more than 200K GZ2 subjects in just 27 days. Our GZX simulation processes a total of \replaced{932,017}{936,887} visual classifications. As presented in Section~\ref{sec: swap is faster}, GZ2 requires at least 35 votes per subject to obtain galaxy classifications that are consistent 95\% of the time. At best, GZ2 could have retired \replaced{26,629}{26,768} subjects with the classifications we process during our GZX run. This implies that we have increased the classification rate by at least a factor of 8, while requiring only 13\% as many human classifications. We next explore the composition of those classifications.

\begin{figure}[t!]
\includegraphics[width=3.35in]{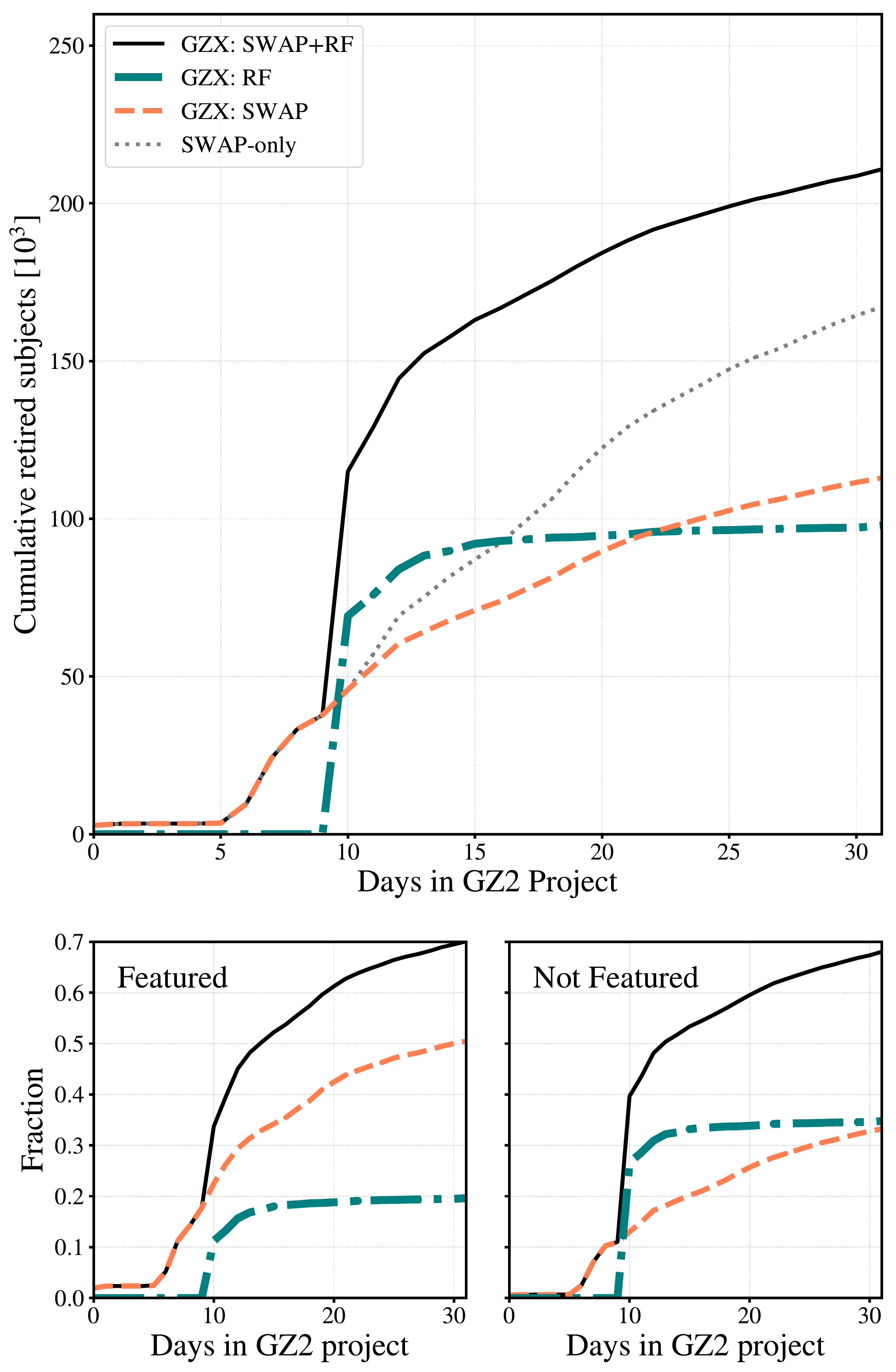}
\caption{Contributions to subject retirement by both classifying agents of GZX: human (SWAP, orange) and machine (RF, teal). The top panel shows cumulative subject retirement for GZX as a whole (solid black), along with that attributed to the RF and SWAP. The dotted grey line shows the fiducial SWAP-only run for comparison. Retirement totals for humans and machine are nearly equal over the course of the simulation but display different behaviours: SWAP's retirement rate is almost constant while the RF contributes substantially after its initial application and then plateaus.  The bottom panels show what fraction of GZ2 subjects are retired, separated by class label. Overall, GZX retires 73.7\% of the entire GZ2 sample in 32 days, retiring the same proportion of~\feat~and~\notfeat~subjects as indicated by the black lines. However, humans retire 30\% more~\feat~subjects than the machine, while both components retire a similar proportion of~\notfeat~subjects. \label{fig: gzx components}}
\end{figure}

\begin{figure*}[t!]
\centering
\includegraphics[width=6.25in]{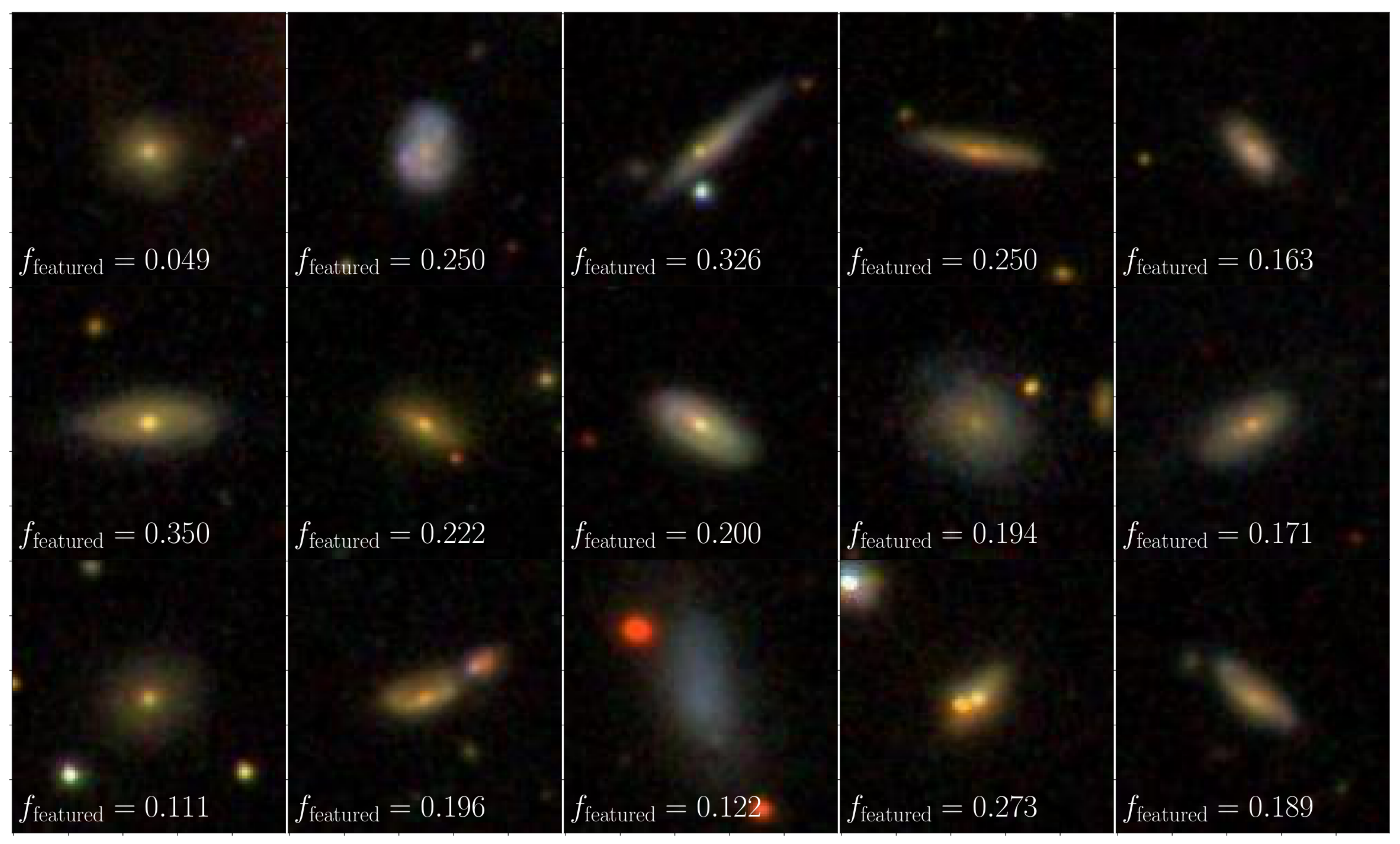}
\caption{A random subsample of subjects identified as false positives: labelled by machine as \feat~but as \notfeat~according to \raw. We display \ffeat~in the lower left corner, that is, the fraction of volunteers who classified the subject as \feat. Values are typically under 0.35 indicating that GZ2 volunteers strongly believed these to be `smooth'  (\notfeat). Fortunately, the machine is able to identify these subjects as~\feat~due to their measured morphology diagnostics.}
\label{fig: machine false pos}
\end{figure*}

\subsection{Who retires what, when?}
\label{sec: who retires what}

In the top panel of Figure~\ref{fig: gzx components} we explore the individual contributions to GZX subject retirement from the RF (dash-dotted teal) and SWAP (dashed orange). The solid black line shows the total GZX retirement (SWAP+RF), while the dotted grey line depicts the fiducial SWAP-only run from Section~\ref{sec: fiducial} for reference. Two things are immediately obvious. First, each component shoulders approximately half of the retirement burden with the machine and SWAP responsible for \replaced{$\sim$$100$K and $\sim$$110$K}{$\sim$$98$K and $\sim$$112$K} subjects respectively. Secondly, the rate of retirement exhibited by the two components is in stark contrast. SWAP retires at a relatively constant rate while the machine retires dramatically at the beginning of its application, quickly surpassing the human contribution, and plateaus thereafter. We thus clearly see three epochs of subject retirement. In the first phase, humans are the only contributors to subject retirement. Once the machine is optimized, it immediately contributes more to retirement than humans. However, the machine's performance plateaus quickly; the third phase is again dominated by human classifications.

In the bottom panels of Figure~\ref{fig: gzx components}, we consider the class
composition of subjects retired by SWAP and the RF. The left (right) panel shows the retired fraction of GZ2 subjects identified as~\feat~(\notfeat) according to their~\raw~labels as a function of GZ2 project time. Overall, GZX retires 73.7\% of the GZ2 subject sample and this is evenly distributed between~\feat~and~\notfeat~subjects as indicated by the solid black lines in both panels. However, SWAP retires more than 50\% of all~\feat~subjects while the machine retires only \replaced{18\%}{20\%}. This divergence does not exist for~\notfeat~subjects where each component contributes \replaced{33-37\%}{33-34\%}. 

What is the source of this discrepancy? Each night the machine trains on a sample composed consistently of 30-40\%~\feat~subjects but does not retire a similar proportion, indicating that the 30\% of non-retired~\feat~subjects do not receive high~\pmachine. In the following section we explore whether this is an artefact of our choice in machine or in the human-machine combination implemented here.

\subsection{Machine performance}
\label{sec: machine performance}

Throughout our analysis we have defined \feat~and \notfeat~subjects by their \raw~labels as this was the most compatible choice for comparison with SWAP output. However, the machine does not learn in the same way, nor is it presented with the same information. Machine and human classifications each provide valuable and complementary information for identifying \feat~galaxies.

We isolate the \replaced{6127}{7060} subjects that were deemed false positives, i.e., galaxies retired by the machine as~\feat~that have~\notfeat~\raw~labels, a sample that comprises only \replaced{6.25\%}{7.2\%} of all subjects the machine retires. We visually examine several hundred and assess that, to the expert eye, a majority are, in fact, \feat. A random sample is shown in Figure~\ref{fig: machine false pos}. 

\begin{figure*}[t!]
\centering
\includegraphics[width=\textwidth]{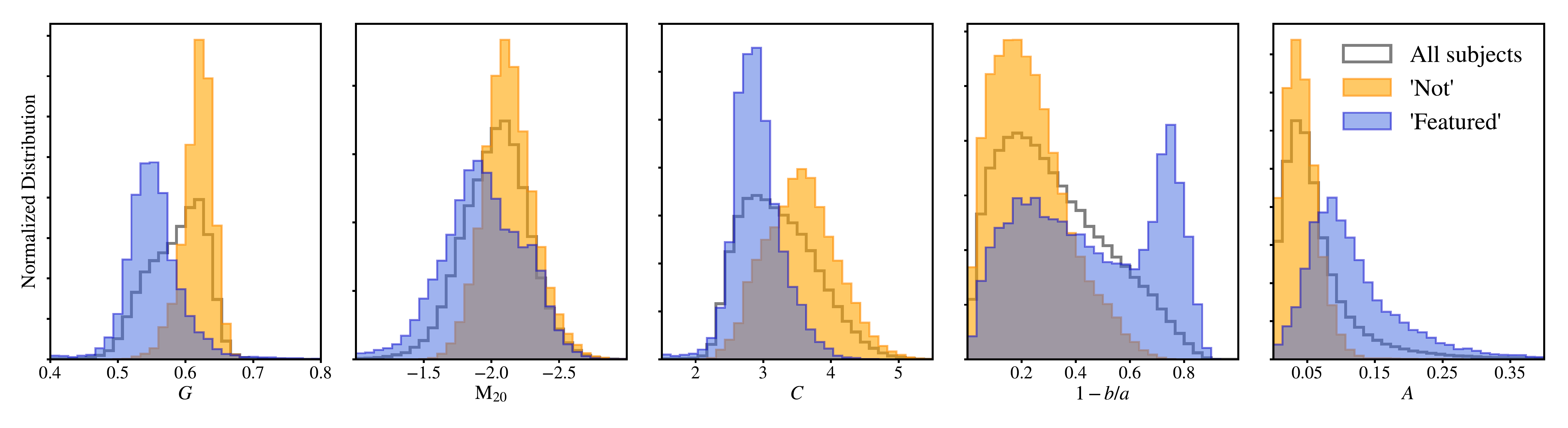}
\caption{The RF is trained on a 5-dimensional morphology parameter space. We show the distribution of each morphology indicator for machine-retired~\feat~(blue) and~\notfeat~(orange) subjects compared to the full GZ2 subject sample (black). The difference between~\feat~and~\notfeat~subjects is in stark contrast for all distributions except, perhaps, \M{20}.  \label{fig: morph params}}
\end{figure*}


That the machine strongly identifies these galaxies as \feat~(\pmachine~$\ge 0.9$) where humans instead classify them as \notfeat~(\ffeat~$< 0.5$) has several contributing factors: 1) as discussed in Section~\ref{sec: swap gz2 disagree}, the threshold we chose carries with it a confidence interval such that subjects with $0.4 <$~\ffeat+\fstar~$< 0.6$ are most likely to receive disagreeing labels from other classifying agents, 2) the first task of the GZ2 decision tree asks a question that does not necessarily correlate with a split between early- and  late-type galaxies, and 3) the machine learns on morphology diagnostics that are very different from visual inspection.

We find that \replaced{41.4\%}{40\%} of these false positives have $0.4 \le$~\ffeat+\fstar$<0.5$ indicating that the disagreement between humans and machine is likely due to the labels we assign at our given threshold. However, we also find that \replaced{43.5\%}{45\%} of false positives have \ffeat+\fstar~$\le0.35$, and this discrepancy is not as easily explained. In Figure~\ref{fig: machine false pos} we examine a random sample of false positives in this regime where, for clarity, we display only the \ffeat~value in the lower left corner. The majority of these subjects are discs lacking features such as spiral arms or strong bars. Whether this is the reason the majority of volunteers classify these objects as ``smooth" is beyond the scope of this paper, however, this behaviour might be modified by providing actual training images and live feedback as performed in \cite{Marshall2016}. We suggest that, at least for this particular question, if either human or machine identifies a subject as \feat, it is likely the subject is discy and worth further investigation.

\begin{figure}[t]
\includegraphics[width=3.2in]{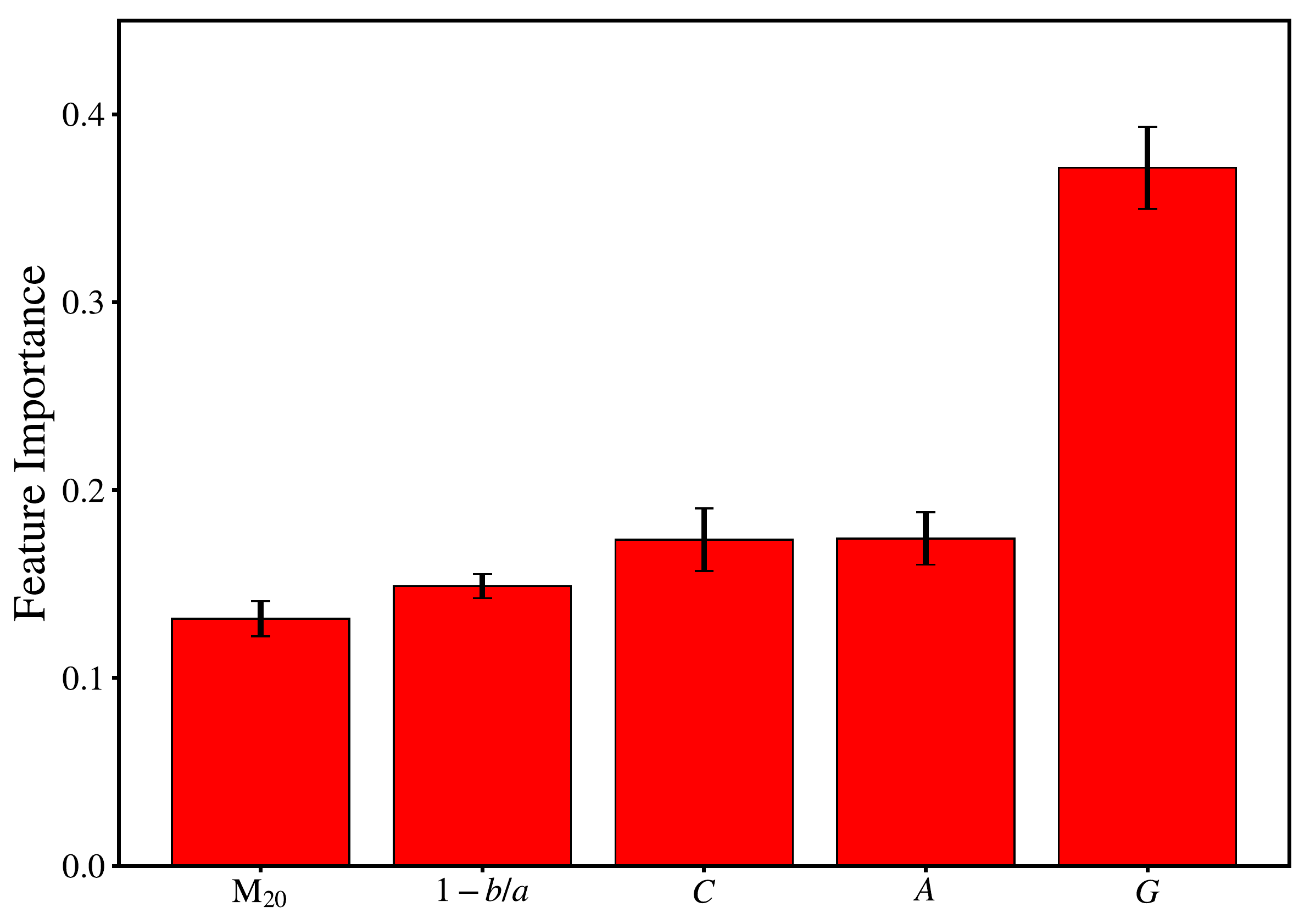}
\caption{The RF's ranked feature importance averaged over all nights of training with black bars indicating the standard deviation. A larger value corresponds to higher importance. The machine computes feature importance according to how much each feature increases the purity of the resulting split averaged over all trees in the forest. The RF places great importance in the Gini coefficient though we note that it can under-represent the importance of highly correlated features such as concentration.\label{fig: feature importance}}
\end{figure}

Accordingly, this suggests that, in some cases, the morphology indicators we measure are sufficient for the machine to recognize~\feat~galaxies regardless of the labels humans provide. Figure~\ref{fig: morph params} shows the distribution of each morphology indicator for all subjects the machine retires as~\feat~(blue) and~\notfeat~(orange) compared to the full GZ2 subject set. The difference between~\feat~and~\notfeat~is stark in all but the \M{20} distribution. This can be seen explicitly in Figure~\ref{fig:  feature importance} in which we show the RF's ranked feature importances, where large values indicate higher importance. Feature importance is computed as how much each feature decreases the impurity of a split in a tree. The impurity decrease from each feature is then averaged over all trees and ranked. We show the feature importance averaged over all nights of training with black bars indicating the standard deviation. The machine finds the Gini coefficient most important for class prediction, placing little emphasis on \M{20}. It is well known that the Gini coefficient is more sensitive to noise than other diagnostics, however, we point out that when a machine is faced with two or more correlated features any of them can be used as the predictor. Once chosen, the importance of the others is reduced. This explains why Concentration is ranked much lower than Gini even though they are strongly correlated as seen in Figure~\ref{fig: morph thresh}. That the machine relies heavily on these two morphology diagnostics is unsurprising as concentration has long been an automated predictor between early- and late-type galaxies~\citep{Abraham1994, Abraham1996, Shen2003}.

The complementary nature of human and machine classification can 
best be utilized by a feedback mechanism in which a portion of machine-retired
subjects are reviewed by humans. Subjects that display excessive disagreement
should be verified by an expert (or expert-user).  In the same way that 
humans increase the machine's training sample over time, subjects that the
machine properly identifies can become part of the humans' training sample.

\section{Looking Forward}
\label{sec: visions}

We have demonstrated the first practical framework for combining human and machine  intelligence in galaxy morphology classification tasks. While we focus below on a brief discussion of our next steps and potential applications to large upcoming surveys, we note that our results have implications for the future of citizen science and Galaxy Zoo in particular. 

GZX is perhaps one of the simplest ways to combine human and machine intelligence and its impressive performance motivates a higher level of sophistication. A first step will be an implementation of SWAP that can handle a complex decision tree. In addition, we envision multiple forms of active feedback in addition to our passive feedback mechanism.  SWAP allows us to leverage the most skilled volunteers to review galaxies difficult for either human or machine to classify.  Additionally, machine-retired subjects should contribute to the training sample for humans in an analogous fashion to what we have already implemented. 

Secondly, our RF can be improved by providing it information equal to what humans receive: multi-band morphology diagnostics will be included in our future feature vector.  However, the Random Forest algorithm is not easily adapted to handle measurement errors or class labels with continuous distributions. A key feature of GZ2 vote fractions is their use in determining the strength of a a morphological feature. Although both SWAP and our RF provide class predictions that are continuous, we apply thresholds to discretize the classification. To fully utilize the information provided, sophisticated algorithms should be considered such as deep convolutional neural networks (CNN) or Latent Dirichlet allocation (LDA), an algorithm that is frequently used in document processing.  Furthermore, there is no reason to limit to a single machine. As hinted at in Figure~\ref{fig: schematic}, several machines could train simultaneously, their predictions aggregated through SWAP, creating an on-the-fly machine ensemble.


With the above upgrades implemented, we expect performance of both the
classification rate and quality to further increase. However, even our current 
implementation can cope with upcoming data volumes from large surveys. 
By some estimates, \textit{Euclid} is expected to obtain measurable morphology with its 
visual instrument (VIS) for approximately $10^6 - 10^7$ galaxies~\citep{Euclid}.
Visual classification at the rate achieved with Galaxy Zoo today
would require 12--120 years to classify.\footnote{We note that the classification 
rate of GZ2 was 4 times higher than GZ's current steady rate.}
If the \textit{Euclid} sample is on the high end, GZX as currently implemented
could classify the brightest 20\% during the six years of its observing mission. 
As currently implemented, we obtain accuracy around 95\% potentially leaving
hundreds of thousands of galaxies with unreliable classifications.  
In a companion paper that seeks to identify supernovae, \cite{Wright2017} 
demonstrate a dramatic increase in accuracy through an entirely different human-machine 
combination whereby the
scores from human and machine are averaged together with the combined score 
yielding the most reliable classification. Again, a combination of both 
approaches will allow us to take full advantage of legacy output from large scale surveys.

\subsection{Conclusions}

In this paper we design and test Galaxy Zoo Express, an innovative system\footnote{Our code can be found at \url{https://github.com/melaniebeck/GZExpress}}for the efficient classification of galaxy morphology tasks that integrates the native ability of the human mind to identify the abstract and novel with machine learning algorithms that provide speed and brute force.  We demonstrate for the first time that the SWAP algorithm, originally developed to identify rare gravitational lenses in the Space Warps project, is robust for use in galaxy morphology classification. We show that by implementing SWAP on GZ2 classification data we can increase the rate of classification by a factor of 4-5, requiring only 90 days of GZ2 project time to classify nearly 80\% of the entire galaxy sample. 

Furthermore, we have implemented and tested a Random Forest algorithm and developed a decision engine that delegates tasks between human and machine.  We show that even this simple machine is capable of providing significant gains in the classification rate when combined with human classifiers: GZX retires over 70\% of GZ2 galaxies in just 32 days of GZ2 project time. This represents a factor of at least 8 increase in the classification rate as well as nearly an order of magnitude reduction in human effort compared to the original GZ2 project. This is achieved without sacrificing the quality of classifications as we maintain $\sim$94\% accuracy throughout our simulations. Additionally, we have shown that training on a 5-dimensional parameter space of traditional non-parametric morphology indicators allows the machine to identify subjects that humans miss, providing  a complementary approach to visual classification. The gain in classification speed allows us to tackle the massive amount of data promised  from large surveys like \textit{LSST} and \textit{Euclid}.

\acknowledgements

\added{We are grateful to the anonymous referees for helpful comments and suggestions which greatly improved this manuscript.} MB thanks Steven Bamford and Boris H{\"a}u{\ss}ler for insightful discussions on citizen science and Galaxy Zoo; and John Wallin and Marc Huertas-Company for several enlightening conversations on machine learning and classification. 
We are grateful to Elisabeth Baeten, Micaela Bagley, Karlen Shahinyan, Vihang Mehta, Steven Bamford, Kevin Schawinski, and Rebecca Smethurst for providing expert classifications in addition to those provided by the authors. PJM acknowledges Aprajita Verma and Anupreeta More for their ongoing collaboration on the Space Warps project. 

MB, CS, LF, KW, and MG gratefully acknowledge partial support from the US National Science Foundation (NSF) Grant AST-1413610. LF and DW also gratefully acknowledge partial support from NSF IIS 1619177.  MB acknowledges additional support 
through New College and Oxford University's Balzan Fellowship as well as the University
of Minnesota Doctoral Dissertation Fellowship. Travel funding was supplied 
to MB, in part, by the University of Minnesota Thesis Research Travel Grant. CJL recognizes support from a grant from the Science \& Technology Facilities Council (ST/N003179/1). 
BDS acknowledges support from Balliol College, Oxford, and the National Aeronautics and Space Administration (NASA) through Einstein Postdoctoral Fellowship Award Number PF5-160143 issued by the Chandra X-ray Observatory Center, which is operated by the Smithsonian Astrophysical Observatory for and on behalf of NASA under contract NAS8-03060. The work of PJM is supported by the U.S. Department of Energy under contract number DE-AC02-76SF00515.
\added{This publication uses data generated via the Zooniverse.org platform, development of which is funded by generous support, including a Global Impact Award from Google, and by a grant from the Alfred P. Sloan Foundation.}

\software{scikit-learn \citep{scikit-learn}, Astropy \citep{astropy}, TOPCAT \citep{topcat}}

\appendix
\label{sec:Appendix}

\section{Exploring the quality of Galaxy Zoo: Express}
\label{sec: vary threshold}

\begin{figure}[t!]
\includegraphics[width=\textwidth]{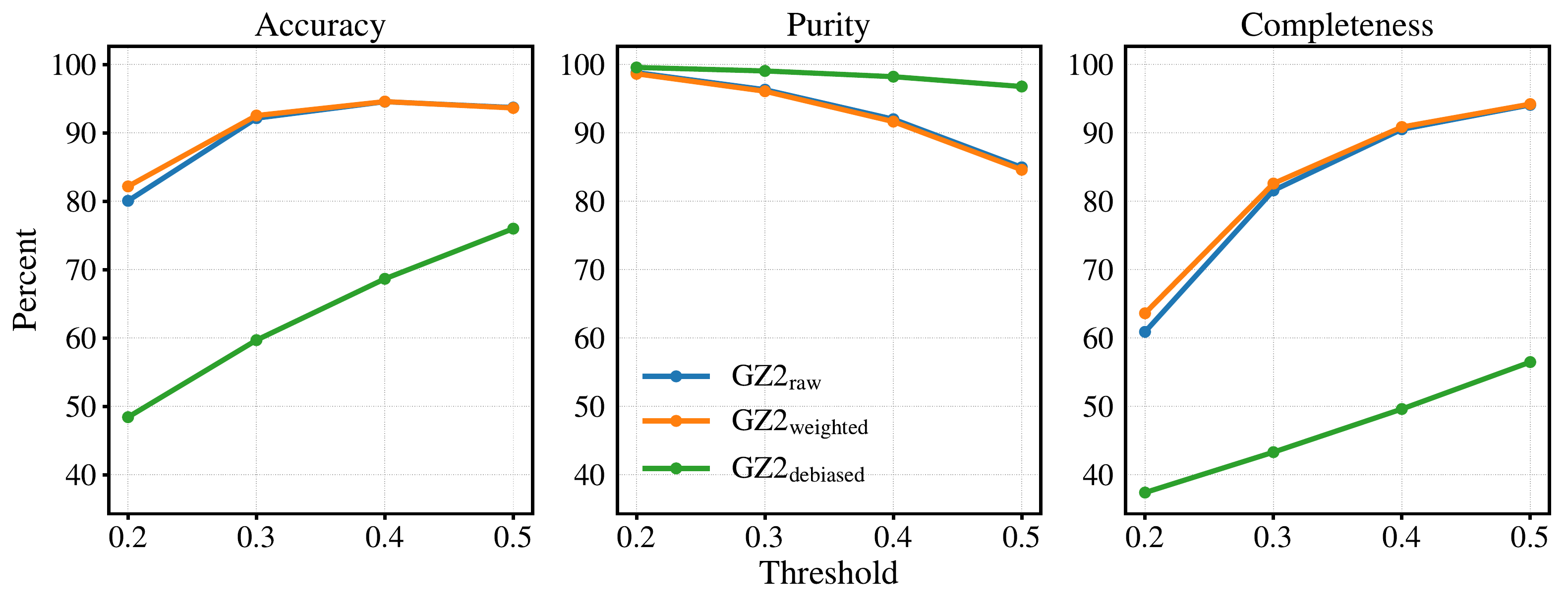}
\caption{Quality metrics computed on the subjects retired during the GZX simulation for a range of thresholds and GZ2 vote fraction types. }
\label{fig: quality}
\end{figure}

In this section we consider the robustness of GZX by computing several sets of ``ground truth'' labels from the GZ2 catalogue. Recall in Section \ref{sec: data} we defined a subject as \feat~if \ffeat+ \fstar $\ge$ \fsmooth, a threshold, $t$, of 0.5. Here we compute new descriptive labels by allowing that threshold to vary where $t \in [0.2, 0.3, 0.4, 0.5]$.  Any subject with \ffeat+\fstar $\ge t$ is labelled \feat, otherwise it is labelled \notfeat. We recalculate the quality metrics of accuracy, purity, and completeness on the sample of galaxies retired during the full GZX simulation (SWAP+RF) for each threshold and each type of GZ2 vote fraction: raw, weighted, and debiased. The results are shown in Figure~\ref{fig: quality}. GZX classifications are quite robust, with accuracy fluctuating by only a few percent for \raw~and \weighted~labels computed using a threshold between 0.3 and 0.5. Instead we see a trade off between purity and completeness. Decreasing the threshold results in more subjects labelled as \feat, which in turn increases sample purity while simultaneously decreasing completeness.

That the \deb~labels perform poorly is not surprising. These vote fractions are computed after considerable post-processing of the raw volunteer votes in order to remove the effects of redshift and surface brightness. As with any set of visual classifications, these biases must be accounted for and this is traditionally done \textit{a posteriori}.  It is also unsurprising that the \raw~and \weighted~classifications are in such tight agreement. \weighted~vote fractions are computed by down-weighting inconsistent volunteers of which there are relatively few. These two sets of vote fractions are thus very similar.

When applying GZX to future imaging programs, there will be no ``ground truth'' labels for comparison. In some sense, these thresholds can be interpreted as a prior for the SWAP \p~value, the initial probability for a subject to be \feat. As we show in Appendix~\ref{sec: tweaking swap}, changing the prior has little affect on the retirement rate but does result in considerable variability in the completeness and purity of the resulting classifications. The choice of whether to optimize SWAP to recover pure or complete samples is a decision for a given science team.

\section{Exploring SWAP's Parameter Space} \label{sec: tweaking swap}

In this Appendix we explore the SWAP parameter space and assess the effects on subject retirement.

\subsection{Initial agent confusion matrix.} 
\begin{figure*}[t]
\includegraphics[width=3.35in]{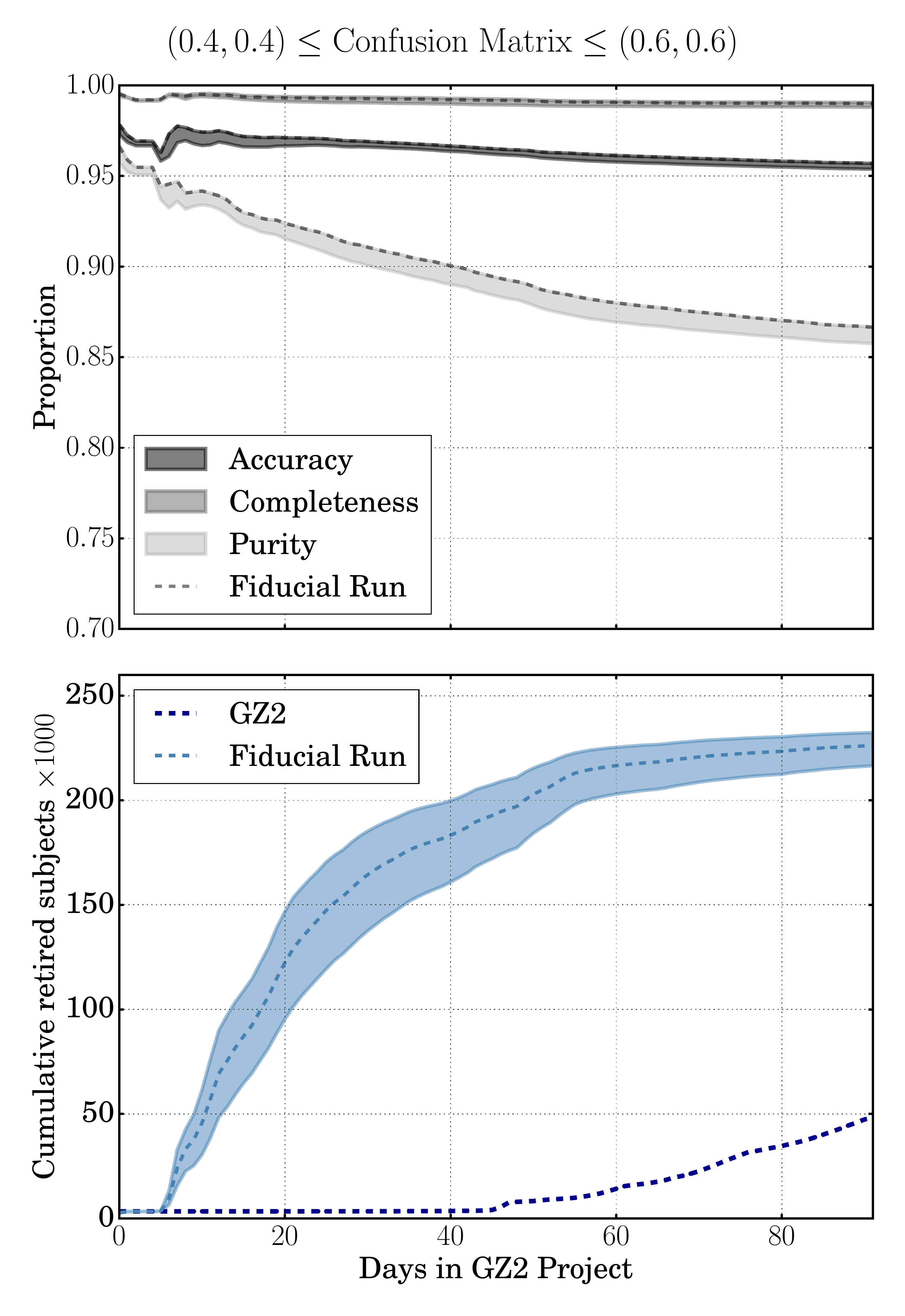}
\includegraphics[width=3.35in]{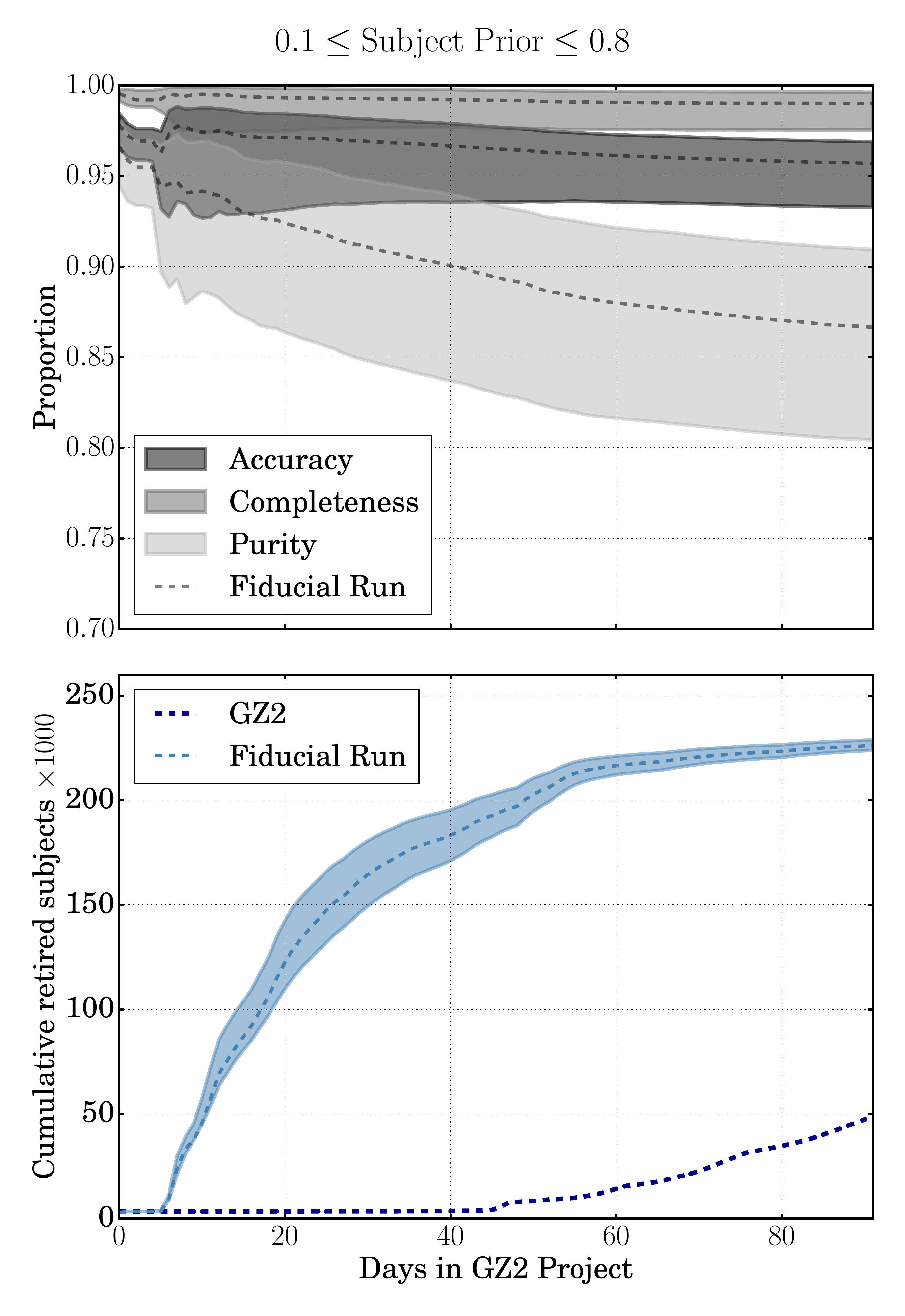}
\caption{SWAP performance does not dramatically change even with a range of input parameters (shaded regions) as compared to the fiducial run of Section~\ref{sec: fiducial} (dashed lines).  \textit{Left.} The quality (top) and retirement rate (bottom) when the confusion matrix is initialized as (0.4, 0.4) and (0.6, 0.6), with all other input parameters remaining constant. \textit{Right.} Same as the left panel but allowing the subject prior probability, \p $= 0.2, 0.35$ and $0.8$. Changing the confusion matrix has little impact on the quality of the labels but varies the total number of subjects retired. In contrast, changing the subject prior is more likely to affect the classification quality rather than the total number of subjects retired. \label{fig: tweak swap}}
\end{figure*}

In our fiducial simulation each volunteer was assigned an agent whose confusion matrix was initialized at $(0.5, 0.5)$, which presumes that volunteers are no better than random classifiers. We perform two simulations wherein we initialize agent confusion matrices as $(0.4, 0.4)$, slightly obtuse volunteers; and $(0.6, 0.6)$, slightly astute volunteers, with everything else remaining constant.  Results of these simulations compared to the fiducial run are shown in the left panel of Figure~\ref{fig: tweak swap}. We find that SWAP is largely insensitive to the initial confusion matrix  both in terms of the subject retirement rate and classification quality.  

We retire $\sim$$225$K$\pm3.5\%$ subjects as shown by the light blue shaded region in the bottom left panel of Figure~\ref{fig: tweak swap}, where the dashed blue line denotes the fiducial run. Predictably, when the confusion matrix probabilities are low, we retire fewer subjects than when these probabilities are high for a given period of time. This is easy to understand since it takes longer for volunteers to become astute classifiers when they are initially given values denoting them as obtuse. Regardless, most volunteers become astute classifiers by the end of the simulation. The top left panel demonstrates our usual quality metrics as computed in Section \ref{sec: fiducial}. The dashed lines again denote the fiducial run. We maintain $\sim$$95\%$ accuracy, $99\%$ completeness, and $\sim$$84\%$ purity;  and no metric changes by $> 2\%$ regardless of initial confusion matrix values.  
 
This spread is due to three effects: 
1) subjects can receive an alternate SWAP label in different simulations, 
2) subjects can be retired in a different order, and 
3) the set of retired subjects is not guaranteed to be common to all runs. 
We find SWAP to be highly consistent: more than 99\% of retired subjects are the same among all simulations, and, of these, 99\% receive the same label.  Instead we find that the order in which subjects are retired changes between runs. When the confusion matrix is low, subjects take longer to classify compared to the fiducial run (i.e., they retire on a later date in GZ2 project time). Likewise, subjects retire sooner when the confusion matrix is high. This can cause quality metrics to vary since they are calculated on a day to day basis. These effects each contribute less than one per cent variation and thus we see a high level of consistency between simulations. 

Of interest, perhaps, is that the quality metrics for these simulations are not symmetric about the fiducial run. However, in the Bayesian framework of SWAP, an agent with confusion matrix (0.4, 0.4) contributes as much information as an agent with confusion matrix (0.6, 0.6). The quality metrics computed are thus within a per cent of each other. In either case, we find that initializing agents at (0.5, 0.5) provides optimal performance for the `training' we simulate with our current approach. Further assessment would require a live project with real-time training and feedback.

\begin{figure*}[t!]
\includegraphics[width=3.08in]{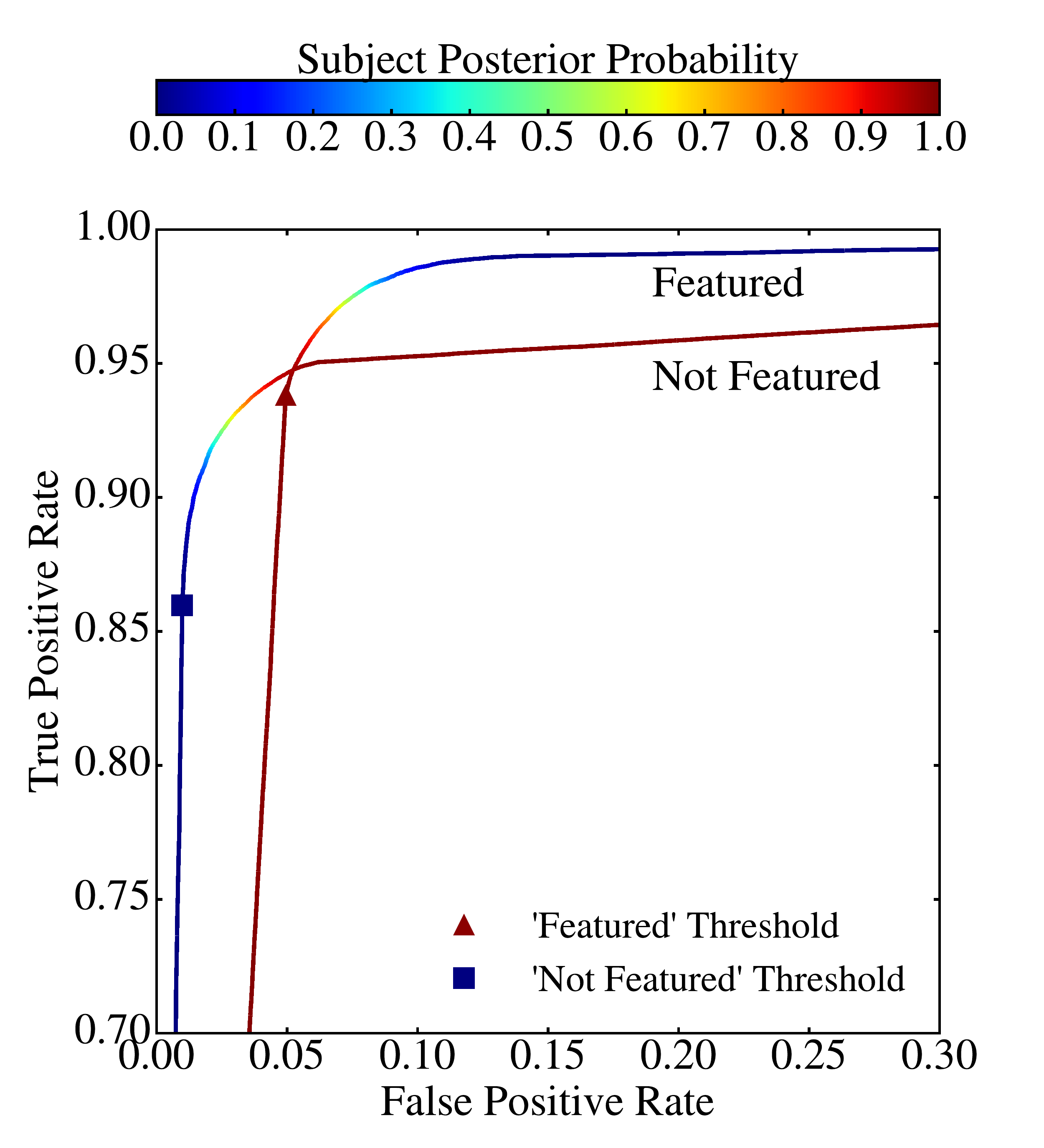}
\includegraphics[width=3.7in]{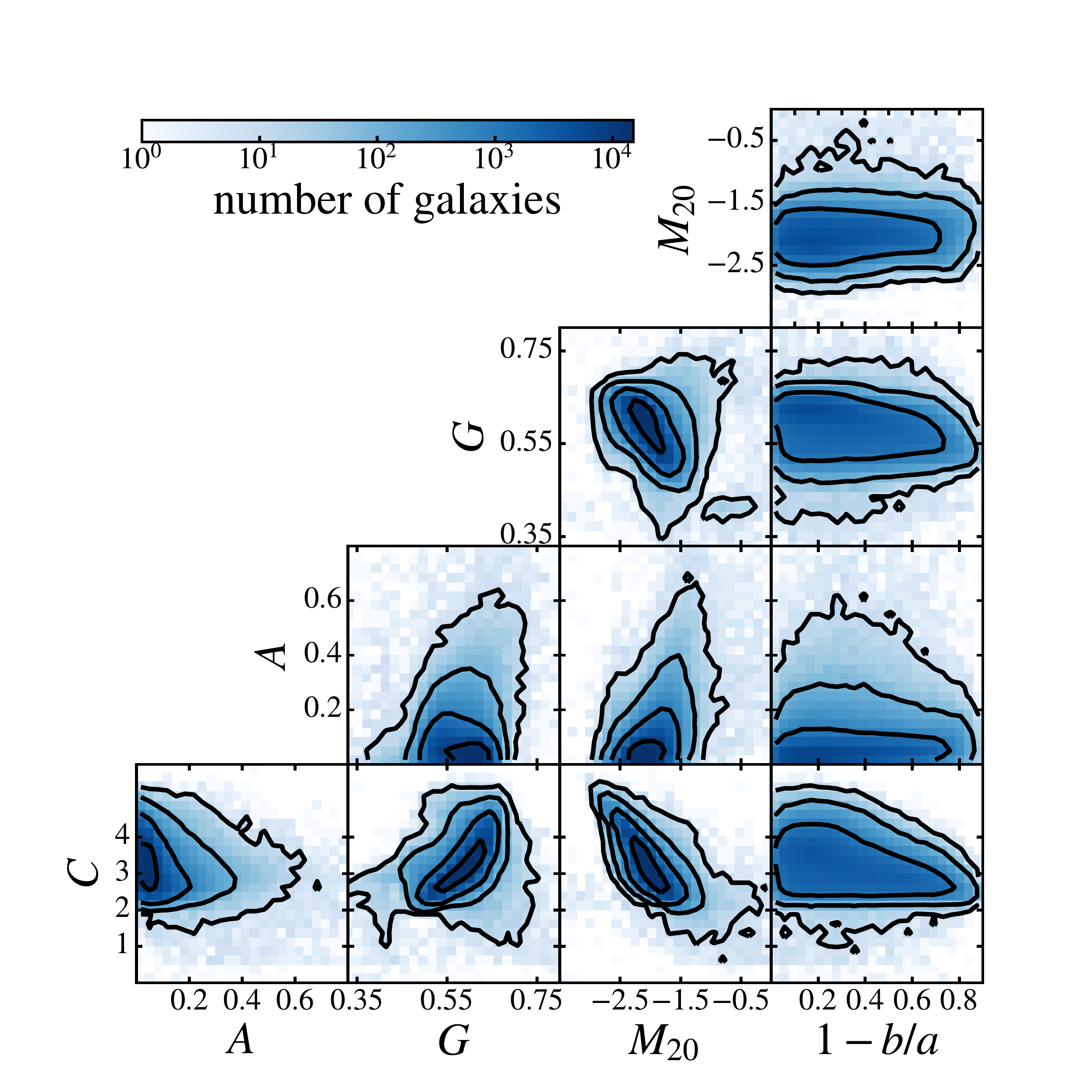}
\caption{\textit{Left.} Identifying~\feat~subjects is independent of identifying~\notfeat~subjects.  Both ROC curves use all subjects processed by SWAP where the score used to create the ROC curve is simply each subject's achieved posterior probability. The Featured curve demonstrates how well we identify~\feat~subjects with a threshold of 0.99, while the Not Featured curve demonstrates how well we identify~\notfeat~subjects with a threshold of 0.004. Typically, best performance is achieved by the score associated with the upper-left-most part of the curve. Our~\feat~threshold is nearly optimal, while our~\notfeat~threshold could be improved since the blue square is not as close to the upper left hand corner as other possible values of the subject posterior. \textit{Right.} Relation between measured morphology diagnostics for more than 280K SDSS galaxies. Most of these galaxies are processed through SWAP, receiving a posterior probability that estimates how likely each is to be~\feat~or~\notfeat.}
\label{fig: morph thresh}
\end{figure*}

\subsection{Subject prior probability,~\p.}
The prior probability assigned to each subject is an educated guess of 
the frequency of that characteristic in the scope of the data at hand. 
For galaxy morphologies, this number should be an estimate of the probability
of observing a desired feature (bar, disk, ring, etc.). In our case, 
we desire simply to find galaxies that are~\feat; however, this is dependent 
on mass, redshift, physical size, etc. The original GZ2 sample was selected
primarily on magnitude and redshift.  As there was no cut on galaxy size
(with the exception that each galaxy be larger than the SDSS PSF), the sample
includes a large range of  masses and sizes. Designating a single prior is not clear-cut; 
we thus explore how various~\p~values effect the SWAP outcome.

We run simulations allowing~\p~to take values 0.2, 0.35, and 0.8 
and compare these to the fiducial run, with everything else remaining constant.
The results are shown in the right panels of Figure~\ref{fig: tweak swap}. 
We again find that SWAP is consistent in terms of subject retirement which varies by only 1\%. 
However, as can be seen in the top panel, the variation in our quality metrics is 
more pronounced. 
Firstly, though we retire nearly the same number of subjects over the course
of each simulation, they are less consistent than our previous runs. That is, 
only 95\% of retired subjects are common to all simulations. Secondly, of those that are 
common, only 94\% receive the same label from SWAP indicating that changing the prior 
is more likely to produce a different label for a given subject than changing the initial 
agent confusion matrix. Finally, there is also a larger spread for the day on which a subject 
is retired as compared to the fiducial run. These trends all contribute to a broader 
spread in accuracy, completeness, and purity as a function of project time.
We stress, however, that although more substantial than the previous comparison, 
these variations are all within $\pm5\%$. 

We can understand these variations more intuitively by considering the following.
Recall that our retirement thresholds,~\tf~and~\tn, have not changed in these simulations. 
When~\p~is small, the subject's probability is already closer to~\tn~in probability space, 
and thus more subjects are classified as~\notfeat~compared to the fiducial run.
Similarly, when~\p~is large, some of these same subjects can instead be classified
as~\feat~because~\p~is already closer to~\tf. Obviously, both outcomes cannot be correct. 
We find that the simulation with~\p~= 0.8 performs the worst of any run; 
this is a direct reflection of the fact that this prior is not suitable for this question or this dataset. 
 Indeed, the best performance is achieved when~\p = 0.35.  This reflects the 
distribution of~\feat~subjects as determined by~\raw~labels and is more characteristic
of the expected proportion of~\feat~galaxies in the local universe.
As a value far from the correct value can have a significant impact on the classification
quality, it is important to choose a prior wisely.

\subsection{Retirement thresholds, \tf~and~\tn.}
Retirement thresholds are directly related to the time that a subject will spend
in SWAP before retirement.  If we lower~\tf~(and/or raise~\tn), more subjects will be retired
compared to the fiducial run as each subject will have a smaller swath of probability space
in which to fluctuate before crossing one of these thresholds.
On the other hand, if we raise~\tf~(and/or lower~\tn), it will take longer for subjects
to cross one of these thresholds. This also increases the likelihood of some subjects 
never crossing either threshold, instead oscillating indefinitely through probability space.

What thresholds should one choose? To answer this question, we consider the left panel of
Figure~\ref{fig: morph thresh}, which depicts the receiver operating 
characteristic (ROC) curve for our fiducial simulation, an illustration of performance as a 
function of a threshold for a binary classifier. 
ROC curves display the true positive rate against the false positive rate for 
a discriminatory threshold or score with a perfect classifier achieving 100\% true positives
and no false positives. The value of the threshold optimal for predicting class labels would 
be that which allows the ROC curve to reach the upper-left-most point in the diagram. 
We have two thresholds to consider and thus we plot the curve twice: 
once under the assumption that ``true positives" denote correctly identified~\feat~subjects; 
and again under the assumption that ``true positives" instead denote correctly identified~\notfeat~subjects.  In both cases, the colour of the line corresponds to the 
subject posterior probability. We mark the location of~\tf~$=0.99$~and~\tn~$=0.004$
from our fiducial run with a red triangle and blue square respectively. 
 We see that~\tf~is nearly optimal but~\tn~could be improved upon.

\section{Measuring Nonparametric Morphological Diagnostics on SDSS Stamps}
\label{sec: measuring morphology}

In order to train our Random Forest machine learning algorithm, we measure  non-parametric morphology diagnostics for the GZ2 galaxy sample. \replaced{We obtain $i$-band imaging from SDSS Data Release 12. Postage stamps are made from the SDSS fields for each galaxy with dimensions of 3 Petrosian radii. Galaxies located within 3 Petrosian radii of the edge of a field were excluded.  Postage stamps undergo a cleaning process whereby nearby sources are identified with SExtractor and their pixels replaced with values that mimic the background in that region.}{We obtain $i$-band imaging (with central wavelength 7480\AA) from SDSS Data Release 12 for 290,059 galaxies, representing 98.2\% of the GZ2 main galaxy sample.  Postage stamps of each galaxy are cut from these fields where the dimensions of each cutout are 4$\times$Petrosian radius as measured by the SDSS pipeline. Galaxies located within 4 Petrosian radii of the edge of a field were excluded as image mosaicking was not performed. This removed 7962 galaxies resulting in a final sample of 282,350 GZ2 galaxy postage stamps, or 95.6\% of the original sample.}

\added{These postage stamps undergo a cleaning process in order to remove the light from nearby sources so as not to contaminate the light profile of the galaxy of interest. Each stamp is processed through Source Extractor \citep[SExtractor, ver. 2.8.6;][]{sextractor}. Two sets of parameters are used as it is not feasible to find a single set of parameters that properly identifies all 282K galaxies. The first is designed to identify bright sources, while the second is better optimized to detect fainter objects. SExtractor segmentation maps are used to  identify the boundaries of each detected object in an image. By design, the galaxy of interest is located at the center of the cutout. Extraneous sources are then identified from both the bright and faint segmentation maps and the pixels corresponding to these sources are replaced with a random value consistent with the background in that postage stamp.}

We compute the following widely adopted nonparametric measurements of the galaxy light distribution on the cleaned postage stamps:

Concentration is computed as $C = 5\log(r_{80}/ r_{20})$ \citep{Bershady2000} where \rr{80} and \rr{20} are the radii containing 80\% and 20\% of the galaxy light respectively.  \added{We define the total flux as that within 1.5 Petrosian radii, and the galaxy center is that determined by the asymmetry minimization \cite[described below,][]{Lotz2004}.} Small values of this ratio tend to indicate discy galaxies, while larger values correlate with early-type ellipticals. 

Asymmetry quantifies the degree of rotational symmetry in the galaxy light distribution (not necessarily the physical shape of the galaxy as this parameter is not highly sensitive to low surface brightness features). A correction for background noise is applied (as in e.g.~\cite{Conselice2000,Lotz2004}), i.e., 
\begin{equation}
A = \frac{\sum_{x,y} |I - I_{180}|}{ 2\sum|I|} - B_{180}
\end{equation}
where $I$ is the galaxy flux in each pixel $(x, y)$, $I_{180}$ is the image rotated by 180 degrees about the galaxy's central pixel, and $B_{180}$ is the average asymmetry of the background. \added{$A$ is summed over all pixels within one Petrosian radius of the galaxy's center and then normalized by a corresponding measure in the original image. The center is determined by minimizing $A$ as described in \cite{Conselice2000}.}

The Gini coefficient, $G$,~\citep{Glasser1962, Abraham2003} describes how uniformly distributed a galaxy's flux is.  If $G$ is 0, the flux is distributed homogeneously among all galaxy pixels; if $G$ is 1,  the light is contained within a single pixel. This term correlates with $C$, however, $G$ does not require that the flux be in the central region of the galaxy.  We follow~\cite{Lotz2004} by first ordering the pixels by increasing flux value, and then computing
\begin{equation}
G = \frac{1}{|\bar X|n(n-1)}\sum_i^n(2i-n-1)|X_i|
\end{equation}
where $n$ is the number of pixels assigned to the galaxy, and $\bar X$ is the mean pixel value. 

\M{20}~\citep{Lotz2004} is the second order moment of the brightest 20\% of the galaxy flux. We compute it as
\begin{eqnarray}
 M_{tot} & = & \sum_i^nf_i[(x_i-x_c)^2 + (y_i-y_c)^2]  \\
 M_{20} & = & \log_{10} (\frac{\sum_iM_i}{M_{tot}}), ~~\textrm{while} \sum_ifi < 0.2f_{tot}
\end{eqnarray}
\replaced{where M$_{tot}$, the total moment, is computed first and $f_{tot}$ is the total flux.}{where $f_i$ is the flux in pixel ($x_i$, $y_i$), and ($x_c$, $y_c$) is the galaxy's center which is determined by minimizing the total moment, $M_{tot}$, in a similar fashion as is done for the asymmetry. The galaxy pixels are then ranked by flux in descending order and $M_i$ is summed over the brightest pixels until that sum equals 20\% of the total galaxy flux within one Petrosian radius, $f_{tot}$, normalized by $M_{\mathrm{tot}}$.}
 For centrally concentrated objects, \M{20} correlates with $C$ but is also sensitive to bright off-centre knots of light. 

Finally, we use the ellipticity, $\epsilon = 1 - b/a$, of the light distribution as measured by SExtractor which computes the semi-major axis $a$ and semi-minor axis $b$ from the second-order moments of the galaxy light.

\begin{table*}
	\centering
	\caption{Summary of morphology measurements made on $\sim$282K galaxies from the GZ2 sample.}

	\label{tab:morph numbers}
	\let\mc\multicolumn
	\begin{tabular}{lccc}
		\mc4c{ \textbf{Morphology measurement summary}} \\
		\hline \hline
			&	&	&	\\
								  & Number &  \% Success & Notes \\
		\hline
		Full Galaxy Zoo 2 sample  	& 295 305 &	   &  \\
		Postage stamps 				& 282 350 &		95.6 	&  \% of full sample\\
		Concentration				& 281 927 &		99.85 	& \% of postage stamps\\ 	
		Asymmetry 					& 282 334 &		99.99 	& \% of postage stamps\\
		Gini coefficient			& 282 323 &		99.99	& \% of postage stamps\\
		M$_{20}$					& 282 194 &		99.94 	& \% of postage stamps\\	
		Ellipticity ($1 - b/a$)		& 282 350 &		100.0 	& \% of postage stamps\\
		\hline
		All morphologies successful & 281 801  &	95.4	& \% of full sample \\
	\end{tabular}
\end{table*}

\replaced{In total, we measure morphological indicators for 282,350 SDSS galaxies.}{In total, we successfully measure all morphological indicators for 281,801 SDSS galaxies. Some galaxies are lost at each stage of the measurement process due to various failures. For example, on rare occasions the minimization of the asymmetry center fails to converge. The number of galaxies with successful measurements at each stage is listed in Table \ref{tab:morph numbers}. } The relations between these diagnostics for the full sample is shown in the right panel of Figure~\ref{fig: morph thresh}. The code developed to clean and compute these morphology indicators is open source and can be found at \url{https://github.com/melaniebeck/measure_morphology}.




\bibliographystyle{apj}
\bibliography{apj-jour,human-machine}


\listofchanges
\end{document}